\documentclass[nolineno]{meltpreprint}
\usepackage{cite}
\usepackage{amsmath,amssymb,amsfonts}
\usepackage{graphicx}
\usepackage{textcomp}
\usepackage{booktabs}
\usepackage{array}
\usepackage{multirow}
\usepackage{url}

\usepackage{bm}
\makeatletter
\AtBeginDocument{\DeclareMathVersion{bold}
\SetSymbolFont{operators}{bold}{T1}{times}{b}{n}
\SetSymbolFont{NewLetters}{bold}{T1}{times}{b}{it}
\SetMathAlphabet{\mathrm}{bold}{T1}{times}{b}{n}
\SetMathAlphabet{\mathit}{bold}{T1}{times}{b}{it}
\SetMathAlphabet{\mathbf}{bold}{T1}{times}{b}{n}
\SetMathAlphabet{\mathtt}{bold}{OT1}{pcr}{b}{n}
\SetSymbolFont{symbols}{bold}{OMS}{cmsy}{b}{n}
\renewcommand\boldmath{\@nomath\boldmath\mathversion{bold}}}
\makeatother

\def\BibTeX{{\rm B\kern-.05em{\sc i\kern-.025em b}\kern-.08em
    T\kern-.1667em\lower.7ex\hbox{E}\kern-.125emX}}

\newcommand{\AiMelt}{57}

\newcommand{\AiStft}{14}
\newcommand{\AmortH}{48}
\newcommand{\AmortV}{26}
\newcommand{\BatchEnHHi}{3.00}
\newcommand{\BatchEnHLo}{1.28}
\newcommand{\BatchEnVHi}{1.45}
\newcommand{\BatchEnVLo}{1.25}
\newcommand{\BatchHHi}{3.08}
\newcommand{\BatchHLo}{1.35}
\newcommand{\BatchVHi}{1.72}
\newcommand{\BatchVLo}{1.46}
\newcommand{\CompiledSaveHHundred}{3}
\newcommand{\CompiledSaveVHundred}{6}

\newcommand{\EnergyAEighteen}{3.03}
\newcommand{\EnergyAgreementMax}{1.0}
\newcommand{\EnergyCompiledHHundred}{2.55}
\newcommand{\EnergyCompiledVHundred}{1.36}
\newcommand{\EnergyHHundred}{2.63}

\newcommand{\EnergyMax}{3.03}

\newcommand{\EquivCcc}{0.9541}
\newcommand{\EquivCosine}{0.9840}
\newcommand{\EquivDelta}{1.06}
\newcommand{\EquivDeltaNat}{0.244}
\newcommand{\EquivRmse}{6.03}

\newcommand{\LatMFourMelt}{0.926}
\newcommand{\LatMFourStft}{2.597}
\newcommand{\MfcctEnAEighteen}{2.22}
\newcommand{\MfcctEnHHundred}{3.65}
\newcommand{\MfcctEnMFour}{2.58}
\newcommand{\MfcctEnVHundred}{1.14}
\newcommand{\MfcctLatAEighteen}{2.55}
\newcommand{\MfcctLatHHundred}{3.54}
\newcommand{\MfcctLatMFour}{2.72}
\newcommand{\MfcctLatVHundred}{1.47}
\newcommand{\MpsgraphMFour}{0.81}
\newcommand{\MsweepEighty}{2.08}
\newcommand{\MsweepHundredTwentyEight}{1.98}
\newcommand{\NnaudioEnRange}{3.8$\times$--9.3$\times$}
\newcommand{\NnaudioLatRange}{4.8$\times$--8.9$\times$}
\newcommand{\PowerAMelt}{3.8}
\newcommand{\PowerAStft}{3.4}
\newcommand{\PowerHMelt}{392.4}
\newcommand{\PowerHRedPct}{20.6}
\newcommand{\PowerHStft}{494.5}

\newcommand{\RidgeV}{17}
\newcommand{\SpeedupAEighteen}{3.29}
\newcommand{\SpeedupCompiledHHundred}{2.27}
\newcommand{\SpeedupCompiledVHundred}{1.86}
\newcommand{\SpeedupHHundred}{2.09}
\newcommand{\SpeedupMFour}{2.80}
\newcommand{\SpeedupMlxAEighteen}{1.28}
\newcommand{\SpeedupMlxMFour}{1.26}
\newcommand{\SpeedupOneAEighteen}{0.74}
\newcommand{\SpeedupOneHHundred}{1.35}
\newcommand{\SpeedupOneMFour}{0.79}

\newcommand{\SpeedupRangeHi}{3.29}
\newcommand{\SpeedupRangeLo}{1.64}
\newcommand{\SpeedupVHundred}{1.64}
\newcommand{\SpiraDropinF}{0.894}
\newcommand{\SpiraDropinPatientF}{0.908}
\newcommand{\SpiraMcnemarB}{7}
\newcommand{\SpiraMcnemarC}{25}
\newcommand{\SpiraMcnemarP}{0.003}
\newcommand{\SpiraMfccF}{0.948}
\newcommand{\SpiraMfccPatientF}{0.973}
\newcommand{\SpiraMfcctF}{0.974}
\newcommand{\SpiraMfcctPatientF}{0.982}
\newcommand{\SpiraOddsRatio}{3.40}
\newcommand{\SpiraPatientCiHi}{+0.050}
\newcommand{\SpiraPatientCiLo}{+0.006}
\newcommand{\SpiraSeedsDeltaPP}{2.7}

\newcommand{\SpiraSeedsP}{0.03125}
\newcommand{\TorchaudioEnRange}{1.5$\times$--2.9$\times$}
\newcommand{\TorchaudioLatRange}{1.8$\times$--4.5$\times$}
\newcommand{\VoxMeltAccMean}{88.3}
\newcommand{\VoxMeltAccStd}{4.5}
\newcommand{\VoxMeltNoninfP}{0.005}
\newcommand{\VoxMfccAccMean}{89.8}
\newcommand{\VoxMfccAccStd}{3.0}
\newcommand{\VoxMfcctAccMean}{89.8}
\newcommand{\VoxMfcctAccStd}{2.2}
\newcommand{\VoxMfcctNoninfP}{0.018}
\newcommand{\VoxStftAccMean}{87.8}
\newcommand{\VoxStftAccStd}{4.5}
\newcommand{\WhisperCiMediumHi}{+0.21}
\newcommand{\WhisperCiMediumLo}{-0.13}
\newcommand{\WhisperCostHi}{2.5}
\newcommand{\WhisperCostLo}{1.7}
\newcommand{\WhisperDwerBase}{+1.67}
\newcommand{\WhisperDwerLarge}{-0.04}
\newcommand{\WhisperDwerMedium}{+0.04}
\newcommand{\WhisperDwerSmall}{+0.63}
\newcommand{\WhisperDwerTiny}{+2.48}
}
\providecommand{\EnergyAgreementMax}{??}
\providecommand{\SpeedupMlxMFour}{??}
\providecommand{\SpeedupMlxAEighteen}{??}

\begin{document}

\title{MelT: A Portable, Single-GEMM Mel Audio Frontend via Non-Uniform DFT with Measured Latency and Energy Gains on GPUs}

\author{\uppercase{Augusto Camargo}\authorrefmark{1}}

\address[1]{Bluecore Consulting, S\~ao Paulo, Brazil (e-mail: augusto.camargo@bluecore.com.br)}

\markboth
{Camargo: MelT: A Portable Single-GEMM Mel Audio Frontend}
{Camargo: MelT: A Portable Single-GEMM Mel Audio Frontend}

\corresp{Corresponding author: Augusto Camargo (e-mail: augusto.camargo@bluecore.com.br).}

\begin{abstract}
Modern neural audio models run on accelerators whose peak throughput comes from dense matrix multiplication, increasingly at the edge and in datacenters. The conventional acoustic frontend, however---a Short-Time Fourier Transform (STFT) followed by sparse Mel aggregation---remains a multi-stage pipeline centered on the Fast Fourier Transform (FFT), with execution overheads unlike the dense linear algebra dominating the inference stack. This work introduces MelT, a portable single-stage Mel frontend that precomputes Mel-spaced Non-Uniform Discrete Fourier Transform (NDFT) bases and applies them to time-domain frames through General Matrix Multiplication (GEMM). The contribution is a computational design principle: decoupling Mel feature extraction from vendor-specific FFT primitives and lowering it onto the matrix-multiplication substrate accelerators already optimize. It is not a new spectral operator. MelT's direct projection performs more arithmetic than the FFT pipeline. Yet in the compact-resolution regime of neural audio frontends, it achieves a \SpeedupRangeLo-times to \SpeedupRangeHi-times latency reduction and up to a \EnergyMax-times reduction in measured active energy, from the Apple A18 Pro to the NVIDIA H100. All gains are within-platform comparisons, accompanied by task-level validation. Word error rate stays statistically equivalent to the native frontend's on frozen Whisper models of medium size and larger; speaker-attribute classification on VoxCeleb1 is non-inferior. The cepstral extension MFCCT preserves utility on a clinical respiratory-insufficiency classification task (SPIRA) while improving on the MFCC baseline. These results indicate that, in practical regimes on the accelerators evaluated here, hardware alignment rather than arithmetic count can govern the realized cost of feature extraction.
\end{abstract}


\titlepgskip=-21pt

\maketitle

\section{Introduction}
\label{sec:introduction}

\begin{figure*}[!t]
\centering
\includegraphics[width=0.88\textwidth]{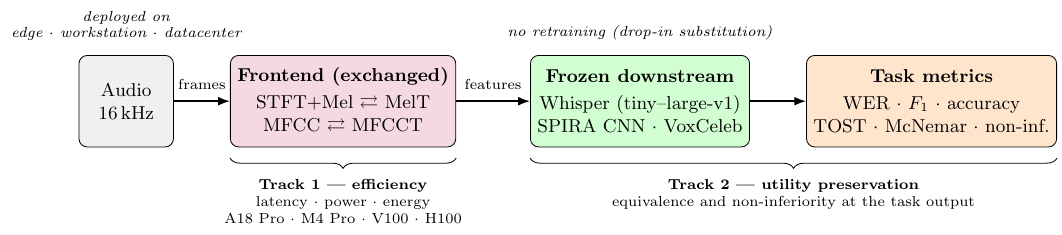}
\caption{\textbf{Experimental design.} The frontend is the only component exchanged. Computational cost (latency, power, and energy) is measured at the frontend boundary on four accelerator platforms (Track~1), while utility preservation is assessed at the task output under frozen downstream models, without retraining (Track~2). Endpoints, datasets, and statistical tests are detailed in Section~\ref{subsec:setup}.}
\label{fig:design}
\end{figure*}

\PARstart{L}{ARGE-SCALE} audio systems---including speech recognizers \cite{radford2022whisper}, audio classifiers \cite{hershey2017cnn,kong2020panns}, and self-supervised models \cite{baevski2020wav2vec}---are increasingly deployed on hardware whose peak throughput is derived from dense matrix multiplication. Modern accelerators, from mobile GPUs to datacenter-scale systems, devote substantial silicon resources to matrix engines such as Tensor Cores \cite{markidis2018nvidia} and specialized matrix coprocessors \cite{jouppi2017datacenter}. As a consequence, many of the dominant workloads in contemporary machine learning are now executed primarily as dense matrix multiplication.

This architectural shift coincides with a change in where inference runs. Foundation models and other AI workloads increasingly run on local and edge hardware---smartphones, personal computers, embedded systems, and accelerator-equipped devices---rather than exclusively in centralized datacenters \cite{zheng2025edgellm,wang2025ondevice}. Vendor platforms reflect this trend: Apple's Foundation Models framework exposes an on-device model to applications \cite{apple_foundationmodels}; Google AI Edge provides an end-to-end stack for on-device generative and multimodal models \cite{google_aiedge}; Microsoft's Foundry Local targets local AI applications and agents \cite{microsoft_foundrylocal}; and NVIDIA supports edge inference through its Jetson platforms \cite{nvidia_jetson}. These efforts are driven by demands for privacy, responsiveness, offline operation, and infrastructure efficiency. As inference spans ever more heterogeneous hardware, latency, energy, and hardware efficiency matter across the entire stack, including the signal-processing frontends that precede neural inference.

The acoustic frontend remains a notable exception. Despite operating within inference pipelines that are otherwise heavily optimized around matrix computation, the dominant frontend architecture continues to rely on a multi-stage pipeline consisting of a Short-Time Fourier Transform (STFT) followed by sparse Mel filterbank aggregation. This architecture reflects the computational assumptions under which it was originally developed---namely the efficiency of the Cooley--Tukey FFT \cite{cooley1965algorithm}---rather than the execution characteristics of modern accelerator hardware. As a result, frontend evaluation may incur additional memory traffic, intermediate tensor materialization, sparse indexing, and kernel-dispatch overheads that are structurally distinct from the dense matrix operations that dominate the remainder of the inference stack.

This observation motivates a broader question: can a conventional audio frontend be reformulated onto the same computational substrate that already underlies modern accelerator execution? The objective is not to introduce a new spectral representation, nor to claim superiority over established Mel-based features. Rather, the objective is to investigate whether a widely used frontend can be decoupled from vendor-specific FFT primitives and re-expressed as dense matrix multiplication while preserving downstream utility.

To explore this question, we revisit the well-established Non-Uniform Discrete Fourier Transform (NDFT) \cite{potts2001fast}. Because the NDFT permits direct evaluation at arbitrary frequency locations, it allows Mel-spaced analysis frequencies to be computed directly rather than obtained through a uniform FFT followed by filterbank aggregation. This direct evaluation enables the frontend to be reformulated as a dense General Matrix Multiplication (GEMM) operation executed on precomputed Mel-spaced basis functions.

We instantiate this design principle as MelT, a single-stage frontend that evaluates Mel-spaced NDFT projections directly through dense matrix multiplication, and MFCCT, its cepstral extension. The claim we make is one of \emph{portability}---the same operator runs on any accelerator exposing a GEMM primitive---rather than of universal performance superiority; the magnitude of any speedup remains hardware-dependent.

Whether such a reformulation is beneficial is not a foregone conclusion. The dense projection performs more arithmetic than the FFT-based pipeline, scaling as $O(NM)$ rather than $O(N \log_2 N)$. A conventional complexity analysis would therefore predict the opposite outcome. The practical question addressed in this work is whether modern matrix-optimized hardware can execute this reformulation more efficiently despite its higher arithmetic cost, and whether any computational gains come at the expense of downstream task performance.

To answer these questions, we evaluate the reformulation across four accelerator platforms spanning mobile-edge, workstation, and datacenter deployments: Apple A18 Pro, Apple M4 Pro, NVIDIA Tesla V100, and NVIDIA H100. Fig.~\ref{fig:design} summarizes the experimental design: computational cost is measured at the frontend boundary, while utility preservation is assessed at the task output under frozen, never-retrained downstream models.

Across these platforms, MelT achieves between \SpeedupRangeLo$\times$ and \SpeedupRangeHi$\times$ latency reduction and up to \EnergyMax$\times$ lower measured active energy. A pre-registered statistical analysis indicates that these computational gains do not require sacrificing utility. Speaker-classification accuracy remains non-inferior, word error rate on medium-sized and larger Whisper models is statistically equivalent under direct frontend substitution, and the cepstral formulation MFCCT demonstrates that the reformulation remains effective on a clinical respiratory-insufficiency classification task.

The primary contributions of this work are summarized as follows:

\begin{itemize}

\item \textbf{A hardware-aligned design principle for audio frontends.}
We show that Mel feature extraction can be decoupled from vendor-specific FFT primitives and reformulated on the matrix-multiplication substrate that already underlies modern accelerator execution.

\item \textbf{A concrete realization of this principle.}
We instantiate the reformulation as MelT and its cepstral extension MFCCT, expressed as Mel-spaced NDFT projections evaluated through dense GEMM operations, and analyze the relationship between direct coherent projection and conventional Mel aggregation.

\item \textbf{A cross-platform computational evaluation.}
We quantify latency, power, and energy behavior across four accelerator platforms spanning edge, workstation, and datacenter deployments, including comparisons against TorchAudio, nnAudio, and librosa.

\item \textbf{Functional validation of the reformulation.}
We demonstrate through equivalence, non-inferiority, and frozen-model substitution experiments that the reformulated frontend preserves downstream utility despite operating through a different computational pathway.

\end{itemize}

Practical operating guidance, including the regimes in which the proposed formulation should not be used, is consolidated in Section~\ref{subsec:whennot}; threats to validity are discussed in Section~\ref{sec:threats}.

\section{Related Work}
\label{sec:related}

This section positions the proposed reformulation within four bodies of
prior work: audio frontend design, dense matrix formulations in signal
processing, the broader systems strategy of casting operations as matrix
multiplication, and energy measurement for AI inference. The organizing
distinction is between methods that seek new audio representations,
methods that accelerate existing frontend implementations, and methods
that change the computational substrate on which an operation is executed.
MelT belongs primarily to the last category.

\subsection{Audio Frontend Design Strategies}

Prior work on audio frontends can be broadly organized into three categories:
(i) learnable frontends that seek improved representations through training,
(ii) accelerated implementations of the conventional frontend pipeline, and
(iii) computational reformulations that reposition signal-processing operations onto accelerator-native execution substrates.
The present work belongs primarily to the third category.

The dominant audio frontend---short-time Fourier transform followed by
triangular Mel filterbank aggregation---was introduced by Davis and
Mermelstein~\cite{davis1980comparison} and has remained architecturally
stable for four decades, despite relying on the Cooley--Tukey
FFT~\cite{cooley1965algorithm}, which was not designed around modern
matrix-accelerator execution paths.

Learnable frontends replace this fixed pipeline with trainable
parameters: SincNet~\cite{ravanelli2018} learns sinc-based filters
from raw waveform; LEAF~\cite{zeghidour2021leaf} extends this approach with
Gabor filters and per-channel energy normalization, at roughly
$300\times$ the compute cost of a fixed Mel spectrogram.
EfficientLEAF~\cite{schluter2022efficientleaf} reduces that overhead
to~$3\%$ of LEAF's cost via inhomogeneous convolutions, yet concludes
that \emph{``both [LEAF and EfficientLEAF] fail to consistently
outperform a fixed mel filterbank---the quest for learnable audio
frontends is not solved''}. Neither learnable approach reports hardware
energy measurements or cross-platform GPU benchmarks.

These approaches pursue a different objective from the present work.
Learnable frontends seek improved representations through training,
whereas the present work investigates whether a fixed and widely deployed
frontend can be reformulated onto a computational substrate that is
better aligned with modern accelerators. The goal is not to replace the
Mel/MFCC representation, but to determine whether that representation can
be executed through a different computational pathway while preserving
its practical utility. A learnable frontend could itself be expressed as
a GEMM-native computation; the contribution here is instead a
reformulation of the fixed frontend already used by production systems.

A parallel line of work accelerates the \emph{conventional} pipeline on
modern hardware rather than reformulating it. nnAudio~\cite{cheuk2020nnaudio},
published in this journal, casts the STFT and the Mel filterbank as
one-dimensional convolution layers, reproducing the standard two-stage
computation on the GPU and benchmarking against \texttt{librosa} by
\texttt{np.allclose} tolerance and visual spectrogram comparison.
TorchAudio~\cite{yang2022torchaudio} similarly provides GPU-resident
STFT and Mel filterbank operators as production building blocks.

Both accelerate the conventional frontend while preserving its underlying
computational structure. The present work instead investigates whether the
frontend itself can be reformulated onto a GEMM-native execution model
(Table~\ref{tab:tmfwc_comparison}).
Where these libraries establish correctness by element-wise tolerance,
the present work additionally quantifies the feature relationship
statistically (Section~\ref{subsec:similarity}) and evaluates
task-level equivalence on a frozen production model
(Section~\ref{subsec:whisper}).

TMFWC~\cite{sebastian2025audio} proposes a time-domain Mel-frequency
representation by synthesizing Mel-spaced sinusoidal basis functions and
computing feature magnitudes directly from time-domain projections.
While conceptually related to the present work in avoiding the
conventional STFT+Mel pipeline, the method is presented using wavelet
terminology and does not explicitly formulate the operation within the
NDFT framework.

More importantly, TMFWC pursues a substantially different objective.
TMFWC does not investigate accelerator alignment, dense GEMM execution,
cross-platform benchmarking, or hardware energy consumption. The present
work instead examines whether Mel-space projection can be reformulated as
a GEMM-native NDFT computation and executed efficiently across
heterogeneous accelerator platforms. Table~\ref{tab:tmfwc_comparison} positions MelT against the closest prior
art on each axis: TMFWC on the formulation axis and nnAudio on the
acceleration axis.

\begin{table}[!t]
\centering
\caption{\textbf{Structural positioning of MelT against the closest prior art on each axis: TMFWC (time-domain Mel formulation) and nnAudio (accelerator-oriented frontend library)}}
\label{tab:tmfwc_comparison}
\setlength{\tabcolsep}{4pt}
\resizebox{\columnwidth}{!}{%
\begin{tabular}{lccc}
\toprule
Property / Feature & TMFWC~\cite{sebastian2025audio} & nnAudio~\cite{cheuk2020nnaudio} & MelT \\
\midrule
Time-Domain Mel Representation        & Yes & No  & Yes \\
Time-Domain Mel Basis Functions       & Yes & No  & Yes \\
Magnitude from Real/Imag.\ Components & Yes & Yes & Yes \\
Wavelet-Theoretic Formulation         & Yes & No  & No  \\
Explicit NDFT Formulation             & No  & No  & Yes \\
Dense GEMM Formulation                & No  & No  & Yes \\
Accelerator-Oriented Design           & No  & Yes & Yes \\
Cross-Vendor Benchmarking             & No  & No  & Yes \\
Measured GPU Energy Telemetry         & No  & No  & Yes \\
\bottomrule
\end{tabular}}
\end{table}
\subsection{Dense Matrix Formulations in Signal Processing}

The Non-Uniform Discrete Fourier Transform (NDFT) has been studied in
signal processing~\cite{bagchi1999book} and geophysical
processing~\cite{duijndam1999}. Unlike the FFT, which evaluates the
spectrum on a uniform frequency grid, the NDFT evaluates the Fourier
transform at arbitrary frequency nodes and can be implemented as a
matrix multiplication against a precomputed basis matrix.

Lin~\cite{lin2018pynufft} formulated the NDFT as a GPU-resident tensor
operation for non-Cartesian MRI reconstruction, reporting
$5$--$13\times$ acceleration over CPU implementations.
Potts et al.~\cite{potts2001fast} provide a widely cited tutorial on
fast algorithms for nonequispaced Fourier transforms.

These works establish that NDFT-as-matmul is computationally viable on
modern accelerators. However, prior work has primarily treated NDFT as a
spectral-analysis tool. A review of the current literature indicates that
prior work has not investigated whether NDFT-based reformulations can
reposition audio frontends onto the matrix-multiplication substrate that
dominates modern accelerator architectures.

\subsection{Casting Operations as Matrix Multiplication}

This pattern is not confined to signal processing; it recurs as a general
systems strategy. Expressing a computation in terms of dense matrix
multiplication to obtain \emph{portable} high performance is a long-established
strategy in high-performance computing. Numerical libraries are deliberately built on
the Level-3 BLAS matrix-multiply kernel~\cite{dongarra1990level3}
precisely so that they run efficiently on any machine that provides an
optimized BLAS, rather than on a hand-tuned kernel per architecture.

Deep-learning systems revived this strategy as accelerators came to
derive their peak throughput from dense matrix multiplication---Tensor
Cores~\cite{markidis2018nvidia} and matrix
coprocessors~\cite{jouppi2017datacenter}. Convolution is the canonical
modern example: the \texttt{im2col} transform reshapes it into a single
matrix multiplication~\cite{chellapilla2006high}, the form used by
production primitives such as cuDNN~\cite{chetlur2014cudnn}. Attention
is likewise realized as batched matrix
multiplication~\cite{vaswani2017attention}.

A complementary line of work instead writes hand-fused kernels---for
example FlashFFTConv~\cite{fu2023flashfftconv} for long FFT-based
convolutions, frequently authored in Triton~\cite{tillet2019triton}.
These kernels can be highly efficient but remain platform-specific,
requiring architecture-dependent engineering effort.

These approaches are instances of a broader systems-design pattern: the
hourglass, or \emph{narrow-waist}, model~\cite{beck2019hourglass}. A
single narrow interface is adopted as a common substrate so that the
operations above it and the hardware below it are decoupled and can
evolve independently. This decoupling is a form of information
hiding~\cite{parnas1972modules}, in which an operation is specified
against a portable interface rather than a particular vendor
implementation.

Dense matrix multiplication is increasingly the narrow waist of the
machine-learning compute stack, with convolution and attention already
expressed above it. Within this landscape the acoustic frontend has
remained a notable exception, still executed through FFT-oriented
spectral primitives rather than dense linear algebra.

The present work does not introduce this strategy but adopts it. Rather
than accelerating a specific FFT implementation, it investigates whether
Mel feature extraction itself can be repositioned above the GEMM narrow
waist. In this formulation, the frontend inherits the portability and
vendor-optimized performance of dense matrix multiplication rather than
depending on a particular FFT library or bespoke platform-specific
kernel.

MelT should therefore be viewed not primarily as a new audio
representation, but as a case study of applying an established
computational strategy---lowering operations onto a portable GEMM
substrate---to the audio frontend.

\subsection{Energy Measurement in AI Inference}

Schwartz et al.~\cite{schwartz2020greenai} argue that computational
efficiency---including energy---should be a first-class evaluation
criterion in AI research alongside accuracy, coining the term
\emph{Green AI}.

Caspart et al.~\cite{caspart2022energy} demonstrate precise GPU energy
measurement via NVML for heterogeneous deep learning workloads,
establishing that runtime alone is insufficient to estimate energy
consumption: actual power draw must be sampled directly from hardware
counters.

This methodology is adopted here---NVML on NVIDIA platforms and
\texttt{powermetrics} on Apple SoCs---and applied to audio frontend
inference, a setting for which hardware-level frontend telemetry is rarely
reported.   

\section{Methodology}

\subsection{Matmul-Native Mel Frontends}
\label{sec:formulations}

This section describes a GEMM-native reformulation of the conventional Mel frontend. Conventional pipelines are multi-stage: a linear-frequency spectrum is first computed and then integrated into the Mel scale through a filterbank. In contrast, Direct Mel Projection evaluates the target Mel-space coordinates directly, avoiding intermediate linear-frequency spectral tensors and expressing the computation as dense matrix multiplication.

Let $x[n]$ denote a discrete-time signal sampled at frequency $f_s$. Mel-spaced analysis frequencies are obtained through the standard Mel mapping

\begin{equation}
\mu(f)
=
2595 \log_{10}
\left(
1 + \frac{f}{700}
\right),
\label{eq:mel_scale}
\end{equation}

with the $M$ analysis frequencies placed at the center frequencies of the triangular Mel filterbank used by conventional pipelines---the HTK Mel convention (\texttt{librosa}~\cite{librosa} with \texttt{htk=True})---uniformly spaced on the Mel scale between $f_{\min}$ and $f_{\max}$. The corresponding analysis frequencies in hertz are obtained through the inverse Mel mapping

\begin{equation}
f_m
=
700
\left(
10^{\mu_m / 2595}
-
1
\right),
\label{eq:inverse_mel}
\end{equation}

where $m = 0,\ldots,M-1$.

Let $x_t[n]$ denote the $t$-th framed signal segment and
$\tilde{x}_t[n] = w[n]x_t[n]$
its windowed version, where $w[n]$ is a symmetric Hann window~\cite{harris1978windows}.

Direct Mel Projection computes real and imaginary projections at the target Mel frequencies as

\begin{equation}
\begin{aligned}
R_{t,m}
&=
\sum_{n=0}^{N-1}
\tilde{x}_t[n]
\cos
\left(
\frac{2\pi f_m n}{f_s}
\right),
\\
I_{t,m}
&=
\sum_{n=0}^{N-1}
\tilde{x}_t[n]
\sin
\left(
\frac{2\pi f_m n}{f_s}
\right).
\end{aligned}
\label{eq:dmp_components}
\end{equation}

The resulting Mel-spaced projection energy is defined as

\begin{equation}
S_{t,m}
=
R_{t,m}^{2}
+
I_{t,m}^{2},
\label{eq:dmp_energy}
\end{equation}

Unlike conventional Mel filtering, in which power values are first computed on uniformly spaced FFT bins and subsequently aggregated through nonnegative triangular weights, Eq.~(\ref{eq:dmp_energy}) performs coherent projection before magnitude-squared energy resolution. Direct Mel Projection should be interpreted as a Mel-spaced NDFT frontend rather than as an algebraically exact rearrangement of the conventional STFT+Mel computation.

The same computation can be expressed natively as dense matrix multiplication.

Let $\mathbf{X}\in\mathbb{R}^{T\times N}$ denote the matrix composed of stacked acoustic frames. Windowing is absorbed into fixed projection matrices defined as

\begin{equation}
\begin{aligned}
\mathbf{W}^{(r)}_{m,n}
&=
w[n]
\cos
\left(
\frac{2\pi f_m n}{f_s}
\right),
\\
\mathbf{W}^{(i)}_{m,n}
&=
w[n]
\sin
\left(
\frac{2\pi f_m n}{f_s}
\right).
\end{aligned}
\label{eq:dmp_weights}
\end{equation}

Because $w[n]$ enters as a fixed per-sample scaling of each basis row, this absorption is mathematically equivalent to windowing each frame before projection and introduces no additional conditioning or numerical-stability penalty. Projection is then performed through two dense matrix multiplications

\begin{equation}
\mathbf{R}
=
\mathbf{X}
\left(
\mathbf{W}^{(r)}
\right)^{\top},
\qquad
\mathbf{I}
=
\mathbf{X}
\left(
\mathbf{W}^{(i)}
\right)^{\top},
\label{eq:dmp_projection}
\end{equation}

followed by element-wise energy resolution

\begin{equation}
\mathbf{S}
=
\mathbf{R}
\odot
\mathbf{R}
+
\mathbf{I}
\odot
\mathbf{I},
\label{eq:dmp_matrix_energy}
\end{equation}

where $\odot$ denotes Hadamard multiplication.

Because $\mathbf{R}$ and $\mathbf{I}$ share the operand $\mathbf{X}$, the two projections of Eq.~(\ref{eq:dmp_projection}) are equivalently a single GEMM against the vertically stacked basis $[\mathbf{W}^{(r)};\,\mathbf{W}^{(i)}]$, whose output is split into its real and imaginary halves. A given backend may execute this as one fused matrix multiplication or as the two separate products, with numerically identical results.

The key computational consequence is that the frontend is reduced to a single dense matrix multiplication followed by element-wise operations. This formulation is independent of FFT libraries and maps directly onto accelerator-native GEMM execution paths.

Subsequent frontend variants differ only in the operations applied to the resulting matrix $\mathbf{S}$.

\subsubsection{MelT Formulation}

MelT (Mel Transform Frontend) applies logarithmic compression directly to the projection-energy matrix:

\begin{equation}
\mathbf{M}^{\mathrm{MelT}}
=
\log
\left(
\mathbf{S}
+
\epsilon
\right).
\label{eq:meltf}
\end{equation}

\subsubsection{MFCCT Formulation}

MFCCT is defined as the cepstral extension of Direct Mel Projection by applying an orthonormal Discrete Cosine Transform (DCT-II)~\cite{ahmed1974dct} matrix
$\mathbf{D}\in\mathbb{R}^{K\times M}$
to the log-compressed representation:

\begin{equation}
\mathbf{C}^{\mathrm{MFCCT}}
=
\mathbf{M}^{\mathrm{MelT}}
\mathbf{D}^{\top}
=
\log
\left(
\mathbf{S}
+
\epsilon
\right)
\mathbf{D}^{\top}.
\label{eq:mfcct_matrix}
\end{equation}

\begin{figure}[!t]
\centering
\includegraphics[width=0.9\columnwidth]{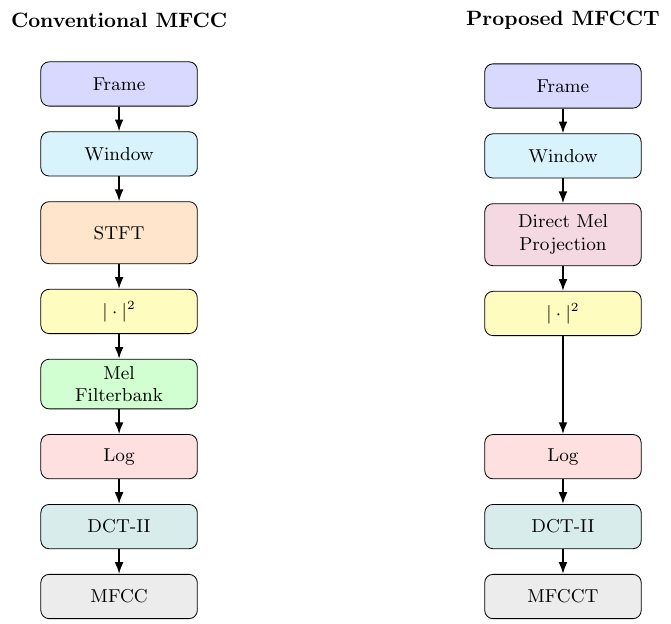}
\caption{\textbf{Computational layout comparing conventional MFCC extraction with the proposed matmul-dense MFCCT frontend configuration.}}
\label{fig:mfcc_mfcct_pipeline}
\end{figure}

The computational distinction between conventional MFCC extraction and the proposed MFCCT formulation is illustrated in Fig.~\ref{fig:mfcc_mfcct_pipeline}.

\subsection{Computational Complexity and Architectural Realization}
\label{subsec:complexity}

The reformulation does not reduce arithmetic complexity; in fact, the opposite is true.

Let $N$ denote frame length, $M$ the number of Mel bins, and $T$ the number of frames. For a single frame, a conventional frontend evaluates an FFT followed by sparse Mel aggregation:

\begin{equation}
\begin{aligned}
\text{Complexity}_{\text{Conv.}}
&=
\mathcal{O}
\left(
N\log_2 N
+
\operatorname{nnz}
\left(
\mathbf{F}_{\mathrm{Mel}}
\right)
\right),
\\
\operatorname{nnz}
\left(
\mathbf{F}_{\mathrm{Mel}}
\right)
&\le
M
\left(
\frac{N}{2}+1
\right).
\end{aligned}
\label{eq:complexity_conv}
\end{equation}

Direct Mel Projection instead evaluates the Mel-spaced NDFT basis directly:

\begin{equation}
\text{Complexity}_{\text{Proposed}}
=
\mathcal{O}
\left(
NM
\right).
\label{eq:complexity_proposed}
\end{equation}

For a complete utterance, both formulations scale linearly with $T$.

From an operation-count perspective, the proposed formulation is therefore less efficient. In the practical regime considered here ($N=400$, $M=80$), the FFT performs substantially fewer arithmetic operations than the direct projection. A purely FLOP-centric analysis would therefore reject the reformulation.

However, realized latency on modern accelerators is frequently dominated by factors beyond operation count, including kernel dispatch overhead, intermediate tensor materialization, memory traffic, sparse indexing, and cache behavior. Reformulating feature extraction as dense General Matrix Multiplication (GEMM) replaces the irregular FFT-plus-filterbank dataflow with a regular compute-bound kernel that maps directly onto vendor-optimized matrix engines.

Rather than relying on vendor-specific FFT implementations, the frontend executes entirely through dense matrix multiplication and element-wise operations. It therefore runs on any accelerator exposing a matrix-multiply primitive---from edge GPUs to datacenter accelerators---without requiring FFT libraries, custom kernels, or platform-specific engineering.

Portability is therefore a property of execution rather than performance: MelT runs wherever dense matrix multiplication is available, while the magnitude of any latency or energy advantage remains dependent on hardware architecture and frontend resolution. Whether this architectural alignment outweighs the higher arithmetic cost on real accelerators is an empirical question, which the following section addresses across four platforms under the two-track design of Fig.~\ref{fig:design}.

\section{Experiments and Results}
\label{sec:experiments}

The following experiments quantify the latency and energy behavior of the
portable single-GEMM formulation across hardware spanning phone-class to
datacenter accelerators. Each result is reported as a within-platform comparison
against the conventional STFT$+$Mel pipeline.

\subsection{Experimental Setup}
\label{subsec:setup}

\subsubsection{Design and Endpoints}
This study is a controlled computational benchmark with a paired
downstream-validation component, organized into the two analysis tracks
summarized in Fig.~\ref{fig:design}, with distinct primary endpoints that are
analyzed separately and never pooled.

The \emph{efficiency} track measures the per-call cost of the proposed front
end against a matched conventional pipeline. Its primary endpoints are the
within-platform latency speedup and the directly-measured active-energy
reduction of MelT relative to STFT$+$Mel at the maximum evaluated context
length (160~s), reported independently on each of the four hardware platforms.
Secondary and sensitivity analyses characterize how those gains scale with
signal duration (1--160~s), Mel-bin count $M$, and batch size, position MelT
against optimized third-party libraries, and isolate the V100 power behavior
under a controlled locked-clock measurement.

The \emph{utility-preservation} track tests whether replacing the conventional
front end with the matmul-native one leaves downstream task performance intact.
Its primary endpoint is the non-inferiority (or equivalence) of task
performance under a frontend swap, evaluated on three independent tasks: word
error rate on a frozen Whisper speech recognizer, respiratory-insufficiency
classification on SPIRA, and speaker-gender classification on VoxCeleb1.
Feature-level fidelity (cosine similarity, concordance, and log-power RMSE) is
reported as a \emph{descriptive} characterization of the representation gap, not
as an equivalence endpoint: MelT and STFT$+$Mel occupy different energy scales
by construction. Equivalence is asserted only at the task level. Table~\ref{tab:analysis_plan} consolidates
this design, mapping each evaluation to its primary endpoint, dataset,
comparator, statistical test, and the table or figure in which its results are
reported.

\begin{table*}[!t]
\centering
\caption{\textbf{Analysis plan.} Each evaluation with its primary endpoint,
dataset and sample size, comparator, statistical test, and pre-declared margin.
The two tracks are reported separately; feature fidelity is descriptive, not an
equivalence endpoint}
\label{tab:analysis_plan}
\setlength{\tabcolsep}{5pt}
\resizebox{\textwidth}{!}{%
\begin{tabular}{llllll}
\toprule
Evaluation & Endpoint & Dataset (size) & Comparator & Test / margin & Reported in \\
\midrule
\multicolumn{6}{l}{\emph{Efficiency track}}\\
Latency        & Speedup @160~s                 & LibriSpeech 1--160~s, 4 platforms        & matched STFT$+$Mel            & bootstrap 90\% CI & Table~\ref{tab:hardware_telemetry}, Fig.~\ref{fig:latency_speedup} \\
Active energy  & Energy reduction @160~s        & same                                     & matched STFT$+$Mel            & bootstrap 90\% CI & Table~\ref{tab:hardware_telemetry}, Fig.~\ref{fig:energy_footprint} \\
External libs  & Latency/energy ratio           & LibriSpeech 100$\times$160~s             & TorchAudio, nnAudio, librosa, \texttt{torch.compile} & bootstrap 90\% CI & Table~\ref{tab:extbaseline}, \S\ref{subsec:extbaseline} \\
Mel-bin / batch & Scaling trend                 & LibriSpeech / Libri6000 4~s              & matched STFT$+$Mel            & point + CI & Figs.~\ref{fig:m_sweep}, \ref{fig:batch} \\
\midrule
\multicolumn{6}{l}{\emph{Utility-preservation track}}\\
Whisper WER    & Equivalence                    & test-clean, 2620 utt / 40 spk            & native log-Mel (frozen)       & TOST (per-speaker), Holm; $\pm1$~pp & Table~\ref{tab:whisper_equiv}, Fig.~\ref{fig:forest} \\
SPIRA          & Non-inferiority / superiority  & 606 win / 108 pat ($\times$6 seeds)      & MFCC                          & McNemar, DeLong, Wilcoxon, bootstrap; $-0.02$ & Tables~\ref{tab:spira_metrics}, \ref{tab:seeds}; Fig.~\ref{fig:forest} \\
VoxCeleb1      & Non-inferiority                & 6 speaker-disjoint splits                & MFCC, STFT$+$Mel              & one-sided NI; $0.01$ & Table~\ref{tab:voxceleb_cross} \\
Feature fidelity & Representation gap (descr.)  & 100 LibriSpeech clips                    & STFT$+$Mel                    & cosine/CCC/RMSE; $\delta{=}\EquivDelta$~dB & Table~\ref{tab:equiv_features} \\
\bottomrule
\end{tabular}}
\end{table*}

\subsubsection{Datasets and Sampling}
\emph{Efficiency track.} Latency, power, and energy are measured on real
LibriSpeech~\cite{panayotov2015librispeech} speech resampled to 16~kHz, with
signal durations from 1 to 160~s. The external-library comparison and the
duration and Mel-bin sweeps use a fixed 100-clip, 160~s subset, and the
batch-scaling sweep uses the Libri6000 corpus of 4~s utterances. Because these
are deterministic timing measurements rather than samples from a population,
the relevant replication is over execution trials rather than over inputs (see
\emph{Statistical Methodology}); a single fixed corpus therefore suffices and is
held constant across platforms and front ends.

\emph{Utility-preservation track.} The Whisper evaluation uses the complete
LibriSpeech test-clean set---2620 utterances from 40 speakers, with no
subsampling. The equivalence test is computed over the entire standard
benchmark, with speakers as the clustering unit (40 clusters). SPIRA's
respiratory-insufficiency classification uses the dataset's held-out test partition (606
analysis windows from 108 patients, balanced 54/54). Because the classifier is
trained from scratch on a small clinical corpus, the test set is augmented with
six independent training seeds for the robustness analysis---the smallest seed
count at which the exact Wilcoxon signed-rank test can reach two-sided
$p<0.05$. VoxCeleb1 gender classification is near its accuracy ceiling and
split-sensitive, so it is evaluated over six speaker-disjoint splits and
reported as mean$\pm$std. The feature-fidelity characterization treats each of
100 LibriSpeech clips as one replicate.

\subsubsection{Comparators and Controls}
The primary comparator throughout is a matched in-house conventional
pipeline---STFT$+$Mel for MelT and MFCC for MFCCT---implemented in the same
framework and configured to the identical DSP specification
(Table~\ref{tab:audio_config}). This matching ensures that every reported speedup, energy
reduction, and downstream comparison isolates the frontend reformulation rather
than an implementation or configuration difference. For the external comparison
(Section~\ref{subsec:extbaseline}) MelT is additionally benchmarked against the
optimized GPU front ends TorchAudio~\cite{yang2022torchaudio} and
nnAudio~\cite{cheuk2020nnaudio} and the CPU reference
\texttt{librosa}~\cite{librosa} (latency only, as no comparable CPU energy
counter is available), each set to the same 80-bin log-Mel specification. We
also benchmark against a compiler-fused conventional baseline
(\texttt{torch.compile}/TorchInductor, CUDA-only) to test whether kernel fusion
alone closes the gap. To control measurement confounds, both front ends in
every paired comparison are run on the same device and backend within a
platform, so device-level factors act on both and largely cancel in the
reported ratios.

\subsubsection{Hardware Infrastructure}
Evaluation was conducted across four execution platforms spanning edge, workstation, and datacenter deployment settings, as summarized in Table~\ref{tab:hardware}.
\begin{table}[!t]
\centering
\caption{\textbf{Hardware platforms used for evaluation}}
\label{tab:hardware}
\setlength{\tabcolsep}{4pt}
\begin{tabular}{llll}
\toprule
Tier & Platform & Memory & Backend \\
\midrule
Edge &
Apple A18 Pro &
Unified &
MPS \\
Workstation &
Apple M4 Pro &
Unified &
MPS \\
Legacy Datacenter &
Tesla V100 &
32 GB &
CUDA + TF32 \\
Modern Datacenter &
H100 &
80 GB HBM3 &
CUDA + TF32 \\
\bottomrule
\end{tabular}
\end{table}
The edge platform is the Apple A18~Pro in the MacBook~Neo, used as a proxy for
smartphone-class on-device inference. Because the iPhone~16 Pro carries the same
A18~Pro silicon, the GPU kernels that execute the front end run on identical
compute units. The headless MacBook also provides the PyTorch/Metal runtime,
corpus access, and \texttt{powermetrics} telemetry that iOS does not expose.
Because the edge metrics are within-device \emph{ratios} measured in a
short-burst regime, device-level factors largely cancel; the validity scope of
this proxy is detailed in Section~\ref{sec:threats}. The workstation platform is
the Mac~mini (2024) with the M4~Pro. NVIDIA results used PyTorch CUDA with TF32,
and both Apple platforms ran both front ends on the PyTorch MPS backend, giving
one consistent backend per platform for all latency, power, and energy
measurements.

\subsubsection{Audio Configuration}
All benchmarks used real LibriSpeech~\cite{panayotov2015librispeech} speech signals resampled to
16~kHz, with durations ranging from
1~s to 160~s. Frontend parameters are summarized
in Table~\ref{tab:audio_config}. Cepstral configurations were selected
to maintain structural symmetry with standard \texttt{librosa}
implementations~\cite{librosa}.

\begin{table}[!t]
\centering
\caption{\textbf{Audio frontend configuration}}
\label{tab:audio_config}
\setlength{\tabcolsep}{4pt}
\begin{tabular}{lcc}
\toprule
Parameter & Spectral & Cepstral \\
\midrule
Sample Rate & 16 kHz & 16 kHz \\
Frame Length $N$ & 400 (25 ms) & 1200 (75 ms) \\
Hop Size $H$ & 160 (10 ms) & 160 (10 ms) \\
Mel Bins $M$ & 80 & 128 \\
MFCC Coefficients $K$ & -- & 40 \\
$f_{\min}$ & 80 Hz & 0 Hz \\
$f_{\max}$ & 7600 Hz & 8000 Hz \\
\bottomrule
\end{tabular}
\end{table}

\subsubsection{Statistical Methodology}
Each (frontend, backend, duration) configuration was measured across 20 independent trials. Each individual trial computed the median of 200 timed execution calls following 50 explicit warmup iterations. Reported metrics reflect the median of the 20 distinct trial medians, providing robustness against transient system interrupt noise. To avoid contention artifacts on these sub-millisecond kernels, the Apple platforms---particularly the passively cooled edge SoC---were freshly booted and measured headless over SSH, with timing started only after background load settled below 5--10\% CPU utilization (interactive GUI sessions, which perturb sub-millisecond timing, were avoided). The NVIDIA datacenter GPUs (H100 and V100) were dedicated exclusively to the benchmark, with no co-resident workloads.

Beyond point estimates, the accompanying reproduction package reports a bootstrap~\cite{efron1979bootstrap} 90\% confidence interval for every headline speedup and for the per-platform power and energy. The controlled V100 power measurement (Section~\ref{sec:discussion}) is also reported as a mean over four repetitions with its standard deviation. The downstream comparisons likewise report effect sizes with significance tests (McNemar~\cite{mcnemar1947note}, DeLong~\cite{delong1988comparing}, Wilcoxon~\cite{wilcoxon1945ranking}) and paired-bootstrap confidence intervals rather than point estimates alone. Multiple-comparison correction is applied within families of tests that bear on a single conclusion. Holm--Bonferroni~\cite{holm1979simple} control is applied to the Whisper equivalence tests across model sizes, since otherwise one could select the model size at which equivalence holds. The VoxCeleb front-end pairs and the multi-seed SPIRA metrics, by contrast, are reported as separate primary and robustness analyses, respectively---each a self-contained result rather than a repeated test of one hypothesis---and are therefore not pooled into a single family. The pre-declared margins are: $\pm1$~percentage-point WER for Whisper equivalence (two one-sided tests, TOST), a $-0.02$ $F_1$ non-inferiority bound for SPIRA, and a 1~percentage-point accuracy margin for VoxCeleb. For feature fidelity, the margin ($\EquivDelta$~dB) is derived from a 1\% $f_{\max}$ retuning (Section~\ref{subsec:similarity}).

\subsubsection{Energy Measurement}
{\sloppy On NVIDIA platforms, per-call energy was measured \emph{directly} from the on-board energy counter (\texttt{nvmlDevice\-Get\-Total\-Energy\-Consumption}, a cumulative whole-board millijoule counter), read at the boundaries of a repeated-call window and divided by the number of completed calls. The window was extended until the counter accumulated at least 30 quantization steps to bound its resolution. On NVIDIA this hardware counter provides an \emph{independent} cross-check against the conventional estimate \(E = \bar{P} \times \bar{t}\) (median sampled power times median per-call latency). The two agree to within \EnergyAgreementMax\%, with the counter no smaller, since it additionally captures the inter-call dispatch and synchronization overhead that \(\bar{t}\) omits. Apple platforms expose no energy counter and instead integrate timestamped GPU power from \texttt{powermetrics} (trapezoidal rule) over the same window. Here both quantities derive from the same power stream, so no independent counter cross-check is available and we report the integrated value. Reported energies are gross (whole-board on NVIDIA, GPU on Apple) and are interpreted as within-platform comparisons between frontend implementations rather than as absolute cross-platform rankings. An idle baseline is captured per platform so net (idle-subtracted) energy can be derived.\par}

\begin{table*}[!t]
\centering
\caption{\textbf{Performance and energy comparison between the conventional STFT+Mel pipeline and the proposed MelT frontend at the maximum evaluated context length (160 s). Metrics use PyTorch CUDA (TF32) on the NVIDIA platforms and the PyTorch MPS backend on the Apple platforms; energy is measured directly per clip (NVML counter on NVIDIA, integrated \texttt{powermetrics} on Apple). Speedup, energy-reduction, and power-reduction factors are computed from full-precision measurements and may differ slightly from ratios of the rounded entries; bracketed values are 90\% confidence intervals}}
\label{tab:hardware_telemetry}
\setlength{\tabcolsep}{6pt}
\resizebox{\textwidth}{!}{
\begin{tabular}{lcccccc}
\toprule
Platform & Latency (ms) & Speedup & Energy (mJ) & Energy & STFT Power (W) & MelT Power (W) \\
 & STFT+Mel / MelT & & STFT+Mel / MelT & Reduction & & \\
\midrule
H100 & $0.146 / 0.070$ & $2.09\times$ & $72.9 / 27.7$ & $2.63\times$ $[2.42, 2.81]$ & $494.5$ $[460.1, 503.9]$ & $392.4$ $[367.7, 394.0]$ \\
V100 & $0.469 / 0.286$ & $1.64\times$ & $110.6 / 76.9$ & $1.44\times$ $[1.42, 1.47]$ & $236.9$ $[236.1, 237.6]$ & $269.7$ $[265.1, 271.5]$ \\
M4 Pro & $2.597 / 0.926$ & $2.80\times$ $[2.79, 2.81]$ & $36.0 / 13.4$ & $2.68\times$ $[2.65, 2.76]$ & $16.5$ $[16.5, 16.6]$ & $16.7$ $[16.5, 16.7]$ \\
A18 Pro & $10.610 / 3.225$ & $3.29\times$ & $28.5 / 9.4$ & $3.03\times$ $[2.92, 3.08]$ & $3.4$ $[3.4, 3.4]$ & $3.8$ $[3.7, 4.1]$ \\
\bottomrule
\end{tabular}
}
\end{table*}

\subsection{Latency and Throughput Analysis}
\label{subsec:latency}

\begin{figure}[!t]
    \centering
    \includegraphics[width=\columnwidth]{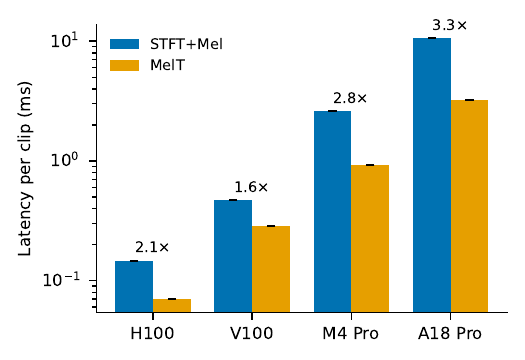}
    \caption{\textbf{Per-clip latency of MelT versus the conventional STFT$+$Mel pipeline at the 160~s context length} (log scale) across the four evaluated platforms. The factor above each pair is the corresponding speedup.}
    \label{fig:latency_speedup}
\end{figure}

Table~\ref{tab:hardware_telemetry} summarizes latency at the maximum evaluated signal duration of 160~s, and Fig.~\ref{fig:latency_speedup} contrasts the two front ends per platform at that duration. Acceleration gains grow with input duration. At the shortest evaluated duration (1~s), backend dispatch overheads dominate. MelT yields a modest $\SpeedupOneHHundred\times$ speedup on the H100 but is slower than the baseline on the Apple devices ($\SpeedupOneMFour\times$ on the M4~Pro and $\SpeedupOneAEighteen\times$ on the A18~Pro), where fixed per-call overhead is not yet amortized over the workload. As audio duration increases to 160~s, the matmul-dense formulation enters a higher-throughput execution regime, establishing latency reductions of \(\SpeedupHHundred\times\) on the H100 and \(\SpeedupAEighteen\times\) on the A18~Pro. These reductions hold despite the projection issuing more arithmetic operations than the FFT pipeline it replaces (Section~\ref{subsec:complexity}). On the Apple M4~Pro workstation, a 160~s signal requires \LatMFourStft~ms with the conventional pipeline, whereas the same segment is processed by MelT in \LatMFourMelt~ms.

On the M4~Pro, the STFT$+$Mel baseline runs faster under Apple's MLX backend than under MPS, so the M4~Pro speedup is backend-dependent. Measured natively under MLX---both front ends reimplemented in pure MLX with no PyTorch---MelT's latency advantage narrows to $\SpeedupMlxMFour\times$ on the M4~Pro ($\SpeedupMlxAEighteen\times$ on the A18~Pro), and the MLX FFT is efficient enough that the two front ends draw comparable energy. We report the MPS configuration ($\SpeedupMFour\times$) throughout because it runs both front ends identically and is the backend instrumented for GPU energy. Implemented natively through Apple's lower-level MPSGraph (again with no PyTorch), MelT processes a 160~s clip in $\MpsgraphMFour$~ms on the M4~Pro, faster than the \LatMFourMelt~ms of the PyTorch/MPS path reported throughout. The PyTorch figures we quote are a conservative bound on MelT's native edge performance.

\subsection{Energy Footprint}
\label{subsec:energy}

\begin{figure}[!t]
    \centering
    \includegraphics[width=\columnwidth]{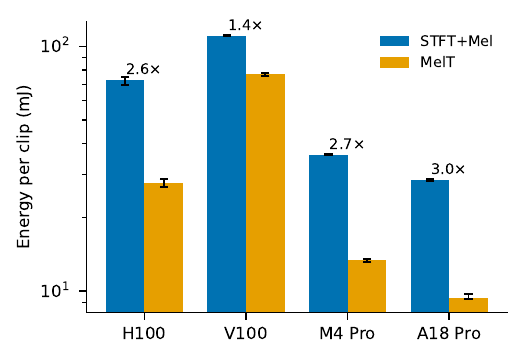}
    \caption{\textbf{Energy per inference at the 160~s context length} (log scale) for MelT versus the conventional STFT$+$Mel pipeline across the four platforms. The factor above each pair is the energy reduction; MelT consumes less energy on every platform.}
    \label{fig:energy_footprint}
\end{figure}

Following the convention of Section~\ref{subsec:setup}, energy is reported as \emph{directly-measured active energy}---the whole-board energy drawn over the active computation window, reported gross (not idle-subtracted)---and as within-platform comparisons between front ends rather than as absolute values. As shown in Table~\ref{tab:hardware_telemetry} and visualized in Fig.~\ref{fig:energy_footprint}, such active-energy reductions can exceed latency gains when lower execution time is accompanied by reduced power draw. On the H100, processing the maximum context yields a median hardware draw of \PowerHStft~W under the conventional STFT$+$Mel pipeline, whereas the dense Direct Mel Projection path yields \PowerHMelt~W. This drop corresponds to a \PowerHRedPct\% reduction in median power draw, which combines with the latency improvement to produce a \(\EnergyHHundred\times\) reduction in active energy. On the A18~Pro, by contrast, MelT draws modestly \emph{higher} instantaneous power (\PowerAStft~W for STFT$+$Mel vs.\ \PowerAMelt~W for MelT). The \(\EnergyAEighteen\times\) active-energy reduction therefore trails the \SpeedupAEighteen$\times$ latency speedup and is attributable to shorter active execution time rather than to lower power.

\subsection{Comparison with External Frontend Libraries}
\label{subsec:extbaseline}

The preceding comparisons isolate the proposed formulation against the matched
in-house STFT$+$Mel reference. To position MelT against optimized third-party
implementations, it was further benchmarked against the external front ends
declared in Section~\ref{subsec:setup}---TorchAudio~\cite{yang2022torchaudio},
nnAudio~\cite{cheuk2020nnaudio}, and the CPU reference
\texttt{librosa}~\cite{librosa}, all at the identical 80-bin log-Mel
specification. Table~\ref{tab:extbaseline} reports MelT's latency speedup and
energy reduction relative to each library.

\begin{table}[!t]
\centering
\caption{\textbf{MelT's latency speedup and energy reduction relative to external libraries (log-Mel, $M{=}80$, 160~s LibriSpeech clips); higher is better. \texttt{librosa} is a CPU reference (latency only).} On the passively-cooled A18~Pro the small MelT energy denominator yields wide upper bounds on the energy-reduction ratios; the lower bounds remain robust}
\label{tab:extbaseline}
\setlength{\tabcolsep}{5pt}
\resizebox{\columnwidth}{!}{
\begin{tabular}{lccccc}
\toprule
 & \multicolumn{2}{c}{TorchAudio} & \multicolumn{2}{c}{nnAudio} & librosa \\
\cmidrule(lr){2-3}\cmidrule(lr){4-5}\cmidrule(lr){6-6}
Platform & Latency & Energy & Latency & Energy & Latency \\
\midrule
H100 & $2.33\times$ & $2.76\times$ $[2.67, 2.90]$ & $7.84\times$ & $9.26\times$ $[9.09, 9.98]$ & $462.42\times$ \\
V100 & $1.85\times$ & $1.51\times$ $[1.45, 1.56]$ & $4.83\times$ & $3.79\times$ $[3.71, 3.92]$ & $337.50\times$ \\
M4 Pro & $2.82\times$ & $2.69\times$ $[2.63, 2.85]$ & $5.84\times$ & $7.55\times$ $[7.13, 8.04]$ & $26.00\times$ \\
A18 Pro & $4.45\times$ & $2.89\times$ $[2.70, 6.34]$ & $8.93\times$ & $7.83\times$ $[7.60, 17.03]$ & $11.76\times$ \\
\bottomrule
\end{tabular}
}
\end{table}

As shown in Table~\ref{tab:extbaseline}, MelT yields the lowest latency and active energy among the evaluated implementations on every platform from mobile (A18 Pro) to
datacenter (H100). It outpaces TorchAudio by \TorchaudioLatRange{} in latency and
\TorchaudioEnRange{} in energy, and nnAudio---the closest prior art (Table~\ref{tab:tmfwc_comparison})---by
\NnaudioLatRange{} in latency and \NnaudioEnRange{} in energy. MelT often draws higher instantaneous power yet uses less total energy by
completing sooner. These third-party libraries are standard production front
ends rather than a fused single kernel. To test whether kernel fusion alone
closes the gap, we additionally compiled the conventional STFT$+$Mel pipeline
end-to-end with \texttt{torch.compile} (the TorchInductor backend, available
only on CUDA) and re-measured it on the H100 and V100. Compiler fusion reduces
the baseline's energy by only \CompiledSaveHHundred--\CompiledSaveVHundred\% and
does not improve its per-call latency on these sub-millisecond kernels; MelT
remains $\SpeedupCompiledHHundred\times$ (H100) and
$\SpeedupCompiledVHundred\times$ (V100) faster, and
$\EnergyCompiledHHundred\times$/$\EnergyCompiledVHundred\times$ lower-energy,
than the compiled baseline. The roofline analysis in
Section~\ref{sec:discussion} accounts for why fusion yields little improvement
here.

\subsection{Representation Fidelity and Downstream Invariance}
\label{subsec:similarity}

As declared in the analysis plan (Section~\ref{subsec:setup}), feature-level
fidelity is reported descriptively and is not an equivalence endpoint. MelT
deliberately reorders projection and energy aggregation, so numerical identity
with STFT$+$Mel is neither expected nor claimed; task-level equivalence is the
pre-registered criterion (Sections~\ref{subsec:whisper}
and~\ref{subsec:spira}). To quantify the downstream impact of this reordering,
frame-level feature matrices were compared using cosine similarity, which
indicates close spatial alignment between MelT and STFT$+$Mel despite the
different order of projection and energy aggregation operations.

To characterize this relationship more rigorously, a pre-registered
equivalence analysis was performed on the H100 over 100 LibriSpeech clips. The
equivalence margin $\delta$ was derived from a parameter change the field
already treats as negligible---a 1\% retuning of $f_{\max}$
($\delta_{\mathrm{RMSE}}{=}\EquivDeltaNat$~nat $=\EquivDelta$~dB). The features are highly correlated in shape (mean per-clip
cosine \EquivCosine, Lin's concordance correlation \EquivCcc) but are \emph{not}
equivalent within this margin: the raw log-power root-mean-square error is
\EquivRmse~dB, exceeding even a 2\% $f_{\max}$ retuning of the same pipeline
(Table~\ref{tab:equiv_features}). We report this
transparently---MelT and STFT$+$Mel sit on different energy scales, so a
TOST on the raw features does not pass. Equivalence is therefore evaluated at the task level rather than on the raw log-power features---after the per-band normalization a network applies, and most directly in end-to-end task performance (Sections~\ref{subsec:whisper} and~\ref{subsec:spira}).

\begin{table}[!t]
\centering
\caption{\textbf{Feature-level comparison of MelT and STFT$+$Mel ($n{=}100$ clips, H100). The raw log-power features are not equivalent within a 1\% $f_{\max}$-derived margin; equivalence is argued at the task level (Section~\ref{subsec:whisper})}}
\label{tab:equiv_features}
\setlength{\tabcolsep}{6pt}
\begin{tabular}{lc}
\toprule
Metric & MelT vs.\ STFT+Mel \\
\midrule
Mean row cosine similarity & $0.9840$ $[0.9837, 0.9843]$ \\
Lin's CCC & $0.9541$ $[0.9531, 0.9551]$ \\
Mean RMSE (dB) & $6.03$ $[6.00, 6.06]$ \\
Anchor margin $\delta$ (dB) & $1.06$ \\
RMSE non-inferior ($<\delta$) & no \\
Cosine non-inferior ($\geq 0.99$) & no \\
\bottomrule
\end{tabular}

\end{table}

\begin{table}[!t]
\centering
\caption{\textbf{Respiratory-insufficiency classification on the SPIRA dataset.} Window- and patient-level metrics for the baseline MFCC and the proposed MFCCT front end on the held-out test set, with McNemar's test, DeLong's test on the ROC-AUC difference, a patient-clustered paired-bootstrap confidence interval, and the drop-in (cross-front-end) result}
\label{tab:spira_metrics}
\setlength{\tabcolsep}{3pt}
\begin{tabular}{lccc}
\toprule
Metric & MFCC & MFCCT & $\Delta$ \\
\midrule
\multicolumn{4}{l}{\textit{Window level}} \\
\quad Accuracy & $0.942$ & $0.972$ & $+0.030$ \\
\quad Precision & $0.911$ & $0.961$ & $+0.050$ \\
\quad Sensitivity & $0.988$ & $0.988$ & $+0.000$ \\
\quad Specificity & $0.891$ & $0.954$ & $+0.063$ \\
\quad $F_1$ & $0.948$ & $0.974$ & $+0.026$ \\
\quad ROC-AUC & $0.992$ & $0.996$ & $+0.004$ \\
\quad PR-AUC & $0.993$ & $0.996$ & $+0.004$ \\
\addlinespace
\multicolumn{4}{l}{\textit{Patient level}} \\
\quad $F_1$ & $0.973$ & $0.982$ & $+0.009$ \\
\quad ROC-AUC & $1.000$ & $1.000$ & $+0.000$ \\
\midrule
\multicolumn{4}{l}{\textit{Statistical tests}} \\
\quad McNemar (window) & \multicolumn{3}{c}{$b{=}7,\,c{=}25,\,p{=}0.003,\,\mathrm{OR}{=}3.40$} \\
\quad DeLong ($\Delta$AUC) & \multicolumn{3}{c}{$p{=}0.0908$} \\
\quad $\Delta F_1$ patient, 90\% CI & \multicolumn{3}{c}{$[+0.006,\,+0.050]$} \\
\quad $\Delta$AUC patient, 90\% CI & \multicolumn{3}{c}{$[-0.002,\,+0.011]$} \\
\quad Drop-in $F_1$ (no retrain) & \multicolumn{3}{c}{window $0.894$ / patient $0.908$} \\
\bottomrule
\end{tabular}

\end{table}

\begin{table}[!t]
\centering
\caption{\textbf{Gender classification on VoxCeleb1 across six speaker-disjoint splits} (accuracy, mean$\pm$std). For each front-end pair the table reports the baseline and proposed same-front-end accuracy, the drop-in (cross-evaluation, no retraining) accuracy, and a one-sided non-inferiority test of the proposed front end against the baseline at a pre-declared 1~percentage-point margin; $p<0.05$ indicates statistical non-inferiority}
\label{tab:voxceleb_cross}
\setlength{\tabcolsep}{4pt}
\begin{tabular}{lcccc}
\toprule
 & \multicolumn{2}{c}{Same front-end} & Drop-in & Non-inf. \\
\cmidrule(lr){2-3}
Pair (base / prop.) & Baseline & Proposed & (no retrain) & $p$ ($\delta{=}1$\,pp) \\
 & acc.\ (\%) & acc.\ (\%) & acc.\ (\%) & \\
\midrule
MFCC / MFCCT & $89.8\!\pm\!3.0$ & $89.8\!\pm\!2.2$ & $88.9\!\pm\!5.6$ & $0.0183$ \\
STFT$+$Mel / MelT & $87.8\!\pm\!4.5$ & $88.3\!\pm\!4.5$ & $86.4\!\pm\!3.5$ & $0.00463$ \\
\bottomrule
\end{tabular}

\end{table}

Downstream evaluation on VoxCeleb1~\cite{voxceleb1} gender classification (Table~\ref{tab:voxceleb_cross}) tests whether the matmul-native front ends preserve task accuracy. This task sits near its accuracy ceiling, with the score dominated by which speakers fall in the held-out fold; the result is assessed as non-inferiority rather than superiority (Section~\ref{subsec:setup}). Both proposed front ends are statistically non-inferior to their baselines---MFCCT versus MFCC ($\VoxMfcctAccMean\pm\VoxMfcctAccStd$ vs.\ $\VoxMfccAccMean\pm\VoxMfccAccStd$, $p{=}\VoxMfcctNoninfP$) and MelT versus STFT$+$Mel ($\VoxMeltAccMean\pm\VoxMeltAccStd$ vs.\ $\VoxStftAccMean\pm\VoxStftAccStd$, $p{=}\VoxMeltNoninfP$)---while the per-split differences are not significant in either direction. Cross-evaluation, in which a classifier trained on the baseline front end is presented the proposed features without retraining, retains most of this accuracy (Table~\ref{tab:voxceleb_cross}, drop-in column).

\subsection{Clinical Respiratory-Insufficiency Classification (SPIRA)}
\label{subsec:spira}

\begin{figure}[!t]
    \centering
    \includegraphics[width=0.95\columnwidth]{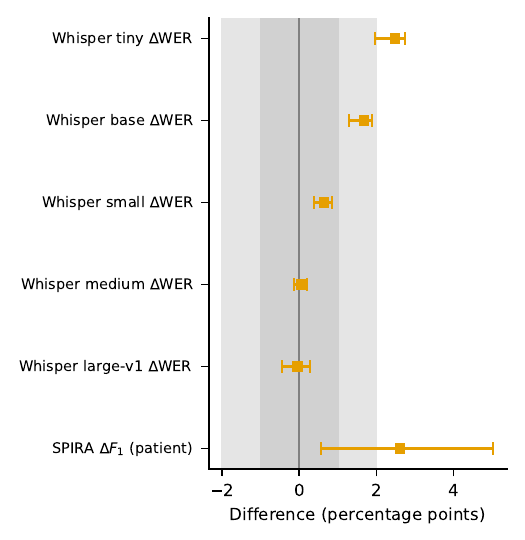}
    \caption{\textbf{Task-level equivalence and non-inferiority.} Per-speaker 90\% confidence intervals for Whisper $\Delta$WER (Section~\ref{subsec:whisper}) and paired-bootstrap intervals for the SPIRA MFCCT$-$MFCC differences, against the $\pm1$ and $\pm2$ percentage-point margins.}
    \label{fig:forest}
\end{figure}

On the clinical respiratory-insufficiency task using SPIRA~\cite{casanova2021spira} COVID-19 classifiers (Table~\ref{tab:spira_metrics}), MFCCT raises window-level $F_1$ from $\SpiraMfccF$ to $\SpiraMfcctF$ and patient-level $F_1$ from $\SpiraMfccPatientF$ to $\SpiraMfcctPatientF$ on the held-out test set. Unlike the near-ceiling speaker-attribute task, this difference is statistically significant. McNemar's test on the discordant window predictions gives $b{=}\SpiraMcnemarB$, $c{=}\SpiraMcnemarC$, $p{=}\SpiraMcnemarP$ (odds ratio $\SpiraOddsRatio$), and a patient-clustered paired bootstrap places the 90\% confidence interval for the $F_1$ difference at $[\SpiraPatientCiLo,\SpiraPatientCiHi]$, entirely above a $-0.02$ non-inferiority bound (Fig.~\ref{fig:forest}). A classifier trained on MFCC and evaluated on MFCCT features without retraining (the drop-in condition) retains an $F_1$ of $\SpiraDropinF$ at the window level and $\SpiraDropinPatientF$ at the patient level (Table~\ref{tab:spira_metrics}).

\subsubsection{Statistical Robustness Across Seeds}
\label{subsec:robustness}
Because the SPIRA classifiers are trained from scratch on a small clinical
corpus, a single training run could in principle favor either front end by
chance. To rule this out, both variants were retrained across six random seeds
and the per-seed differences compared with a Wilcoxon signed-rank test
(Table~\ref{tab:seeds}). MFCCT exceeds MFCC on all six seeds for window $F_1$,
window ROC-AUC, and patient ROC-AUC ($p{=}\SpiraSeedsP$ in each case---the smallest
two-sided value attainable at this sample size). It exceeds MFCC on five of the six seeds
for patient $F_1$, with a mean window-$F_1$ advantage of $+\SpiraSeedsDeltaPP$~percentage
points. The difference is therefore consistent across seeds.

\begin{table}[!t]
\centering
\caption{\textbf{SPIRA multi-seed robustness.} Mean$\pm$std across six training seeds, with a Wilcoxon signed-rank test on the per-seed MFCCT$-$MFCC differences}
\label{tab:seeds}
\setlength{\tabcolsep}{5pt}
\resizebox{\columnwidth}{!}{
\begin{tabular}{lcccc}
\toprule
Metric & MFCC & MFCCT & $\Delta$ & Wilcoxon $p$ \\
\midrule
$F_1$ (window) & $0.9457\!\pm\!0.0218$ & $0.9730\!\pm\!0.0081$ & $+0.0273$ & $0.03125$ \\
ROC-AUC (window) & $0.9882\!\pm\!0.0069$ & $0.9984\!\pm\!0.0015$ & $+0.0102$ & $0.03125$ \\
$F_1$ (patient) & $0.9604\!\pm\!0.0227$ & $0.9805\!\pm\!0.0143$ & $+0.0201$ & $0.0625$ \\
ROC-AUC (patient) & $0.9978\!\pm\!0.0019$ & $0.9999\!\pm\!0.0001$ & $+0.0021$ & $0.03125$ \\
\bottomrule
\end{tabular}
}
\end{table}

\subsection{Functional Equivalence on a Frozen Whisper Model}
\label{subsec:whisper}

The cross-evaluation and equivalence analyses above operate at the feature
and small-classifier level. To test whether MelT's features are
interchangeable with a conventional front end on a large production model
(the drop-in substitution of Track~2 in Fig.~\ref{fig:design}),
the log-Mel front end of OpenAI's Whisper~\cite{radford2022whisper} was
replaced with MelT and transcription quality measured on the full
LibriSpeech test-clean set. The Whisper
weights are frozen and unmodified; only the front end is swapped, and
decoding is greedy, so any change in word error rate (WER) is attributable
solely to the front end. MelT is configured to Whisper's 80-bin log-Mel
specification; this is an \emph{alternative} GEMM-native front end, not a
reimplementation of Whisper's filterbank---its feature-space cosine to
Whisper's native mel is only 0.84.

\begin{table}[!t]
\centering
\caption{\textbf{MelT as a drop-in front end for frozen Whisper (LibriSpeech test-clean, 2620 utterances, 40 speakers). Decoding is greedy, so $\Delta$WER isolates the front end. Equivalence is assessed by two one-sided tests (TOST) on the per-speaker $\Delta$WER against $\pm1$ and $\pm2$ percentage-point (pp) margins}}
\label{tab:whisper_equiv}
\setlength{\tabcolsep}{4pt}
\resizebox{\columnwidth}{!}{
\begin{tabular}{lccccc}
\toprule
Model & WER native & WER MelT & $\Delta$WER (pp) & 90\% CI (pp) & TOST $\pm1$/$\pm2$ \\
\midrule
tiny & 9.66 & 12.13 & +2.48 & $[1.965,2.749]$ & -- / -- \\
base & 6.84 & 8.51 & +1.67 & $[1.285,1.886]$ & -- / \checkmark \\
small & 4.93 & 5.56 & +0.63 & $[0.386,0.845]$ & \checkmark / \checkmark \\
medium & 4.48 & 4.52 & +0.04 & $[-0.131,0.209]$ & \checkmark / \checkmark \\
large-v1 & 4.20 & 4.16 & -0.04 & $[-0.451,0.275]$ & \checkmark / \checkmark \\
\bottomrule
\end{tabular}
}
\end{table}

As shown in Table~\ref{tab:whisper_equiv}, the WER is statistically
equivalent to the native front end from the
\emph{medium} model upward despite the low feature-space similarity. On
Whisper-medium, $\Delta$WER is $\WhisperDwerMedium$~pp with a per-speaker 90\% confidence
interval of $[\WhisperCiMediumLo,\WhisperCiMediumHi]$~pp, and on large-v1 MelT is marginally better
($\WhisperDwerLarge$~pp); both pass a TOST of equivalence at a strict $\pm1$~pp margin.
Equivalence is assessed at the speaker level, since utterances are not
independent within a speaker; the per-utterance test agrees. The residual gap
shrinks monotonically with model capacity ($\WhisperDwerTiny \to \WhisperDwerBase \to \WhisperDwerSmall \to
\WhisperDwerMedium \to \WhisperDwerLarge$~pp from tiny to large-v1). Smaller models (tiny, base) are
\emph{not} equivalent---the front-end change costs \WhisperCostLo--\WhisperCostHi~pp---whereas
higher-capacity models absorb the substitution within the tested equivalence margins, consistent with the
two front ends differing mainly by a per-band, affine-like transformation to which an
expressive model becomes largely invariant. Functional equivalence thus holds at the
task level precisely where it fails at the representation level, with the
threshold set by model capacity.

\subsection{Computational Generalization to MFCCT}
\label{subsec:mfcct_compute}

Because MFCCT reuses the same GEMM-native Direct Mel Projection stage employed by MelT and differs only through the addition of logarithmic compression and a fixed DCT-II transform, similar computational behavior is expected. To verify that the observed hardware benefits are not specific to the MelT representation, MFCC and MFCCT were benchmarked on the same hardware platforms and evaluation protocol.

\begin{table}[!t]
\centering
\caption{\textbf{Maximum observed acceleration and energy reduction of MFCCT relative to conventional MFCC extraction at 160 s}}
\label{tab:mfcct_summary}
\setlength{\tabcolsep}{8pt}
\begin{tabular}{lcc}
\toprule
Platform & Latency Gain & Energy Gain \\
\midrule
H100 & $3.54\times$ & $3.65\times$ $[3.53, 3.91]$ \\
V100 & $1.47\times$ & $1.14\times$ $[1.13, 1.14]$ \\
M4 Pro & $2.72\times$ & $2.58\times$ $[2.48, 2.65]$ \\
A18 Pro & $2.55\times$ & $2.22\times$ $[2.19, 2.25]$ \\
\bottomrule
\end{tabular}

\end{table}

As shown in Table~\ref{tab:mfcct_summary}, the cepstral extension preserves the computational advantages of the proposed formulation across all evaluated hardware platforms, achieving up to a $\MfcctLatHHundred\times$ latency reduction and a $\MfcctEnHHundred\times$ energy reduction on the H100 relative to conventional MFCC extraction. Gains remain substantial on Apple Silicon (M4~Pro $\MfcctLatMFour\times$/$\MfcctEnMFour\times$; A18~Pro $\MfcctLatAEighteen\times$/$\MfcctEnAEighteen\times$) and are smallest on the V100 ($\MfcctLatVHundred\times$/$\MfcctEnVHundred\times$). Because MFCCT uses a longer analysis frame and a higher Mel-bin count than MelT, its per-platform ordering differs from the MelT results of Table~\ref{tab:hardware_telemetry}; nonetheless, the benefits of matmul-native frontend computation clearly extend beyond MelT and persist after cepstral transformation.

\subsection{Scaling with the Number of Mel Bins}
\label{subsec:m_sweep}

To empirically characterize the scaling behavior implied by the \(\mathcal{O}(N M)\) projection cost, an additional sweep was performed over the number of Mel bins \(M\) on the NVIDIA H100 at the maximum evaluated context length of 160~s.

\begin{figure}[!t]
\centering
\includegraphics[width=0.95\columnwidth]{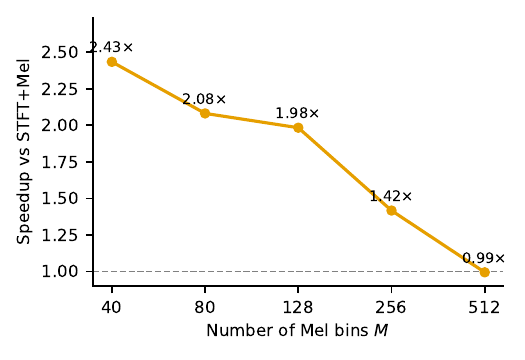}
\caption{\textbf{MelT speedup over the conventional STFT$+$Mel pipeline as a function of the number of Mel bins \(M\)} (NVIDIA H100, 160~s). The advantage decreases monotonically with \(M\), consistent with the \(\mathcal{O}(NM)\) projection cost, and approaches break-even near \(M{=}512\).}
\label{fig:m_sweep}
\end{figure}

As shown in Fig.~\ref{fig:m_sweep}, the speedup decreases
monotonically as \(M\) increases, consistent with the expected
\(\mathcal{O}(N M)\) arithmetic scaling of Direct Mel Projection.
Nevertheless, the proposed formulation remains advantageous
throughout the \(40\)--\(512\) Mel-bin range evaluated in this
study. In particular, MelT retains speedups of
\(\MsweepEighty\times\) and \(\MsweepHundredTwentyEight\times\) at \(M=80\) and \(M=128\),
respectively, corresponding to the operating regime commonly
used by contemporary neural audio frontends. The results indicate that the practical crossover point occurs near the upper end of the evaluated range, while MelT remains advantageous in the Mel-resolution regime typically employed in modern speech and audio systems.

\subsection{Throughput Scaling with Batch Size}
\label{subsec:batch}

\begin{figure}[!t]
\centering
\includegraphics[width=0.95\columnwidth]{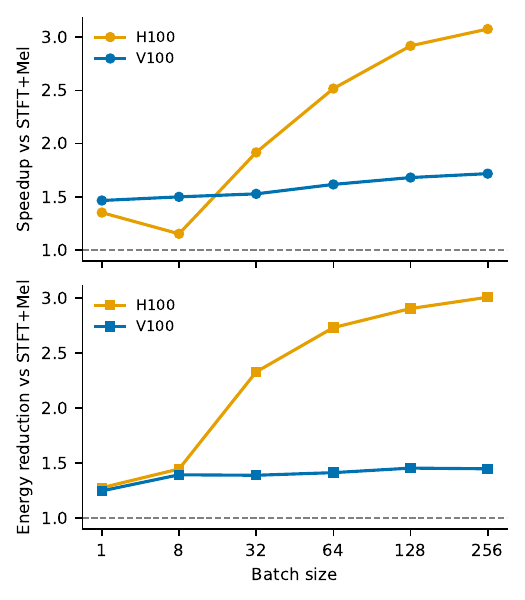}
\caption{\textbf{MelT speedup (top) and active-energy reduction (bottom) over the conventional STFT$+$Mel pipeline as a function of batch size} (Libri6000, 4~s utterances; GPU compute time, host-to-device transfer measured separately; per-batch energy from the on-board counters). On the H100 both grow with batch size as larger batches saturate the GEMM units; on the V100 they grow more modestly. MelT is faster and lower-energy than STFT$+$Mel at every batch size on both platforms.}
\label{fig:batch}
\end{figure}

Datacenter inference is typically served in batches rather than one
utterance at a time, so we additionally characterize throughput as a function
of batch size. Using the Libri6000 corpus partitioned into full batches of
4~s utterances (50 timed runs and 10 warmups per batch size, GPU compute time;
host-to-device transfer measured separately), we sweep the batch size from
1 to 256 on the NVIDIA H100 and V100. Results are shown in
Fig.~\ref{fig:batch}.

On the H100, MelT's speedup over the conventional pipeline grows
from $\BatchHLo\times$ at batch~1 to $\BatchHHi\times$ at batch~256, as
larger batches more fully saturate the GEMM units; on the V100 it grows more
modestly, from $\BatchVLo\times$ to $\BatchVHi\times$ across batch sizes. In both cases MelT
remains faster than STFT$+$Mel at every batch size. Per-utterance latency is
strongly amortized by batching---roughly $\AmortH\times$ on the H100 and
$\AmortV\times$ on the V100 from batch~1 to batch~256. Active energy follows the
same trend. The per-batch energy reduction of MelT over the conventional
pipeline grows from $\BatchEnHLo\times$ at batch~1 to $\BatchEnHHi\times$ at
batch~256 on the H100, and from $\BatchEnVLo\times$ to $\BatchEnVHi\times$ on
the V100 (measured per batch from the same on-board counters;
Section~\ref{subsec:setup}). The matmul-native frontend therefore benefits from the same
batching that accelerator-based serving already exploits, with the relative
advantage increasing on the compute-rich H100.

\subsection{Data and Code Availability}

Our reproduction package---the MelT/MFCCT implementation, all experiment
runners, configuration files, per-platform launch scripts, and the aggregated
results that generate every table and figure---is publicly available at
\url{https://github.com/augustocamargo/MelT}, with an archived release on
Zenodo (DOI:~10.5281/zenodo.20635529). A manifest maps each table and figure to the exact command and output
that produces it, and a preflight check verifies a machine's readiness before
any run.

The datasets used are all publicly available. LibriSpeech~\cite{panayotov2015librispeech}
(test-clean, CC-BY~4.0) is obtained from OpenSLR. The fixed-length Libri6000
corpus used for the latency, energy, and equivalence measurements is
reconstructed byte-for-byte from LibriSpeech train-clean-100 (also OpenSLR,
CC-BY~4.0) by the provided script, from a recorded source manifest with
SHA-256 checksums---no audio is redistributed.
VoxCeleb1~\cite{voxceleb1} is obtained from the VoxCeleb project under its
research license. The SPIRA dataset~\cite{casanova2021spira} is publicly
available under a CC~BY-SA~4.0 license at
\url{https://github.com/SPIRA-COVID19/SPIRA-ACL2021}. Because VoxCeleb1
and SPIRA carry licenses that restrict or condition redistribution, the package
links to their official sources rather than re-hosting them; the
LibriSpeech-derived corpus is reconstructed locally from its CC-BY~4.0 source
rather than redistributed.

\textit{Ethics statement.} The SPIRA corpus~\cite{casanova2021spira} was
collected anonymously, with no personally identifying or demographic
information, and released publicly under a CC~BY-SA~4.0 license. The present
study performs a secondary analysis of this public, de-identified dataset and
involved no new data collection from human subjects.

\section{Discussion}
\label{sec:discussion}

These results recast the acoustic frontend as a hardware-alignment problem rather than a transform-design one. The question introduced in Section~\ref{sec:introduction} was whether a conventional audio frontend could be repositioned onto the same dense matrix-multiplication substrate that already underlies modern accelerator execution. Sections~\ref{sec:related} and \ref{sec:formulations} argued that such reformulations are a well-established systems strategy, while Section~\ref{sec:experiments} quantified their practical consequences. MelT and MFCCT should be viewed as concrete instantiations of a broader design principle: decoupling frontend computation from vendor-specific FFT primitives and expressing it directly through dense GEMM operations.

Within contemporary machine-learning systems, dense matrix multiplication increasingly acts as a computational narrow waist. Convolution, attention, and most large-scale neural workloads are already expressed above this substrate, yet the conventional audio frontend remains unusual in continuing to depend on FFT-specific implementations.

Portability, in particular, is a property of the formulation rather than of any platform. Because the frontend reduces to matrix multiplication and element-wise operations, it can execute on any platform exposing a GEMM primitive without requiring FFT libraries, custom kernels, or architecture-specific engineering. The CUDA, MPS, MLX, and MPSGraph implementations reported in Section~\ref{sec:experiments} provide direct evidence of this property. MelT need not outperform every FFT implementation on every device; it removes the frontend's dependence on FFT-specific infrastructure while staying on the execution model that modern accelerators already optimize most aggressively.

The latency results are consistent with this accelerator-native perspective. Although Direct Mel Projection performs more arithmetic operations than the FFT-based pipeline, it executes through a dense compute-bound kernel rather than through a sequence of FFT, aggregation, indexing, and element-wise stages. The observed gains arise not from reduced operation count, but from a more favorable execution model on matrix-oriented accelerators.

MelT reduced execution time on every evaluated platform, from \SpeedupVHundred$\times$ on the NVIDIA V100 to \SpeedupAEighteen$\times$ on the Apple A18 Pro. The spread tracks the backend-specific factors discussed in Section~\ref{subsec:complexity} rather than asymptotic complexity alone. The largest gains occur where matrix-oriented execution paths are particularly efficient and where the cost of intermediate tensor materialization and irregular memory access is highest. In this sense, the results reinforce a broader observation increasingly visible throughout modern computing: realized performance is often governed more strongly by execution substrate than by arithmetic count alone.

The energy measurements reveal two distinct regimes. On the H100, the energy reduction exceeds the latency gain because shorter execution time is accompanied by a \PowerHRedPct\% reduction in power draw. On the A18 Pro, MelT draws modestly higher instantaneous power (\PowerAMelt~W versus \PowerAStft~W), so the energy reduction is attributable primarily to shorter execution time. In both cases, however, active energy decreases because the computation completes sooner. Since energy per inference---the quantity that governs battery drain---is lower and each inference is a sub-millisecond burst, the higher instantaneous power carries no adverse thermal or battery implication on the edge device.

\begin{figure}[!t]
    \centering
    \includegraphics[width=0.9\columnwidth]{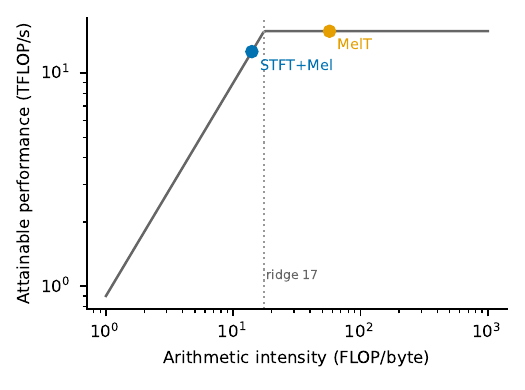}
    \caption{\textbf{Roofline analysis of the V100 power inversion.} MelT (single GEMM, $\approx$\AiMelt~FLOP/byte) is compute-bound while STFT$+$Mel ($\approx$\AiStft~FLOP/byte) is memory-bound, sitting on opposite sides of the V100 compute/bandwidth ridge ($\approx$\RidgeV~FLOP/byte). MelT draws higher instantaneous power yet completes sooner, so its energy per clip is lower (Table~\ref{tab:hardware_telemetry}).}
    \label{fig:roofline}
\end{figure}

The Tesla V100 provides the clearest illustration of the architectural tradeoff. MelT executes as a single dense GEMM and is compute-bound, whereas STFT$+$Mel remains memory-bound. The roofline analysis in Fig.~\ref{fig:roofline} places MelT at approximately \AiMelt~FLOP/byte and STFT$+$Mel at approximately \AiStft~FLOP/byte, on opposite sides of the V100 compute/bandwidth ridge ($\approx$\RidgeV~FLOP/byte). Consequently, MelT draws higher instantaneous power while achieving lower total energy.

The computational advantages are also not specific to MelT. Its cepstral extension MFCCT reuses the same Direct Mel Projection stage, adding only logarithmic compression and a fixed DCT-II transform, yet preserves the same latency and energy trends across all evaluated platforms (Table~\ref{tab:mfcct_summary}). The gains arise from the matmul-native projection stage itself rather than from a particular output representation.

The external-library comparison further supports this interpretation. MelT remains faster and lower-energy than TorchAudio, nnAudio, and the compiled STFT$+$Mel baseline despite these implementations already representing highly optimized realizations of the conventional pipeline. Compiler fusion reduces dispatch overhead, but it does not fundamentally alter the bandwidth-bound character of the FFT-centered computation. The primary advantage comes from changing the computational substrate rather than from incremental implementation-level optimization.

Taken together, the latency, energy, roofline, and external-baseline results suggest that the observed behavior is not an artifact of a particular implementation. Rather, it emerges from a structural difference between FFT-centered and GEMM-centered execution models. This interpretation is consistent with the broader systems literature reviewed in Section~\ref{sec:related}, where computational reformulation has repeatedly been used to align algorithms with dominant hardware primitives.

From a signal-processing perspective, Direct Mel Projection should not be interpreted as an algebraically exact reformulation of conventional Mel filtering. These two pipelines differ in the ordering of projection and energy aggregation, producing a coherent Mel-spaced projection rather than an incoherent aggregation of FFT-bin energies. The relevant question, however, is not whether the representations are identical, but whether the reformulated frontend remains useful.

The evidence suggests that it does. Feature-level equivalence is not achieved (Section~\ref{subsec:similarity}), yet task-level utility is preserved across every downstream evaluation. The two front ends differ mainly by a per-band affine transformation---a scale and offset on each Mel channel---that input normalization and the first learned layer absorb, which is why task-level performance is robust even where the raw representations diverge. VoxCeleb results demonstrate statistical non-inferiority, frozen-model Whisper substitution demonstrates functional equivalence for medium-sized and larger models, and SPIRA demonstrates that the reformulated cepstral frontend remains highly effective on a clinical respiratory-insufficiency classification task. In this sense, downstream utility survives the computational reformulation even when feature-level equivalence does not.

The SPIRA results deserve particular care in interpretation. MFCCT exceeds the MFCC baseline on this task, and the improvement is statistically significant. However, the purpose of the experiment is not to establish representational superiority in general. Rather, it demonstrates that the reformulated frontend preserves utility and, in this application, yields higher performance than the conventional formulation. The paper argues preservation of utility as the primary claim and treats performance improvements as task-specific observations.

The latency and energy gains take on added relevance as foundation-model inference moves onto heterogeneous local hardware, where preprocessing enters the operational inference budget rather than remaining a negligible implementation detail \cite{murshed2021edge}. In that setting, a frontend expressible on any matrix-multiplication device---without a vendor-specific FFT dependency---is portable precisely where deployment is most fragmented. The significance of this work also extends beyond Mel features themselves: other frontends that can be expressed as fixed linear transforms and are currently realized through FFT-centric execution may be lowered onto the same matrix-multiplication substrate. No empirical claim is made here regarding such frontends; rather, MelT is presented as a fully evaluated case study demonstrating that matrix-oriented frontend reformulation is feasible, portable, and practically beneficial on modern accelerators.

\subsection{Practical Guidance}
\label{subsec:whennot}
The results reduce to an operating envelope. MelT is the appropriate choice in three situations:
\begin{itemize}
  \item \textbf{Compact spectral resolution ($M\approx64$--$128$).} The configuration range that dominates deployed neural audio systems; the measured advantage persists through $M{=}128$ (Fig.~\ref{fig:m_sweep}).
  \item \textbf{Long contexts or batched serving.} The advantage grows with audio duration on every platform and compounds under batching, where per-utterance amortization raises both throughput and per-batch energy reduction (Section~\ref{subsec:batch}).
  \item \textbf{Heterogeneous deployment targets.} Because the operator requires only a GEMM primitive, a single formulation serves CUDA, MPS, and MLX backends without FFT-library dependence---useful when one codebase must span datacenter and edge.
\end{itemize}

Conversely, the same results delineate where MelT is \emph{not} the appropriate choice:
\begin{itemize}
  \item \textbf{High spectral resolutions ($M>512$).} The $\mathcal{O}(NM)$ projection reaches break-even with the FFT pipeline near $M{=}512$ (Fig.~\ref{fig:m_sweep}) and is asymptotically unfavorable beyond it; the conventional STFT$+$Mel pipeline remains preferable for high-resolution analysis.
  \item \textbf{Small frozen models requiring exact feature compatibility.} Low-capacity models (e.g., Whisper tiny/base) do not absorb the feature-scale difference under zero-shot substitution, incurring a $\WhisperCostLo$--$\WhisperCostHi$~pp WER penalty; such models should be retrained on MelT features, or paired with a calibration step, rather than swapped in frozen.
  \item \textbf{Applications demanding algebraic equivalence to a Mel filterbank.} MelT is a coherent, not algebraically exact, reformulation (\EquivRmse~dB feature RMSE; Section~\ref{subsec:similarity}); use cases that require bit-faithful reproduction of conventional Mel energies should retain the standard pipeline.
  \item \textbf{Very short, single-clip edge workloads.} At the shortest durations on passively cooled edge hardware, fixed per-call overhead is not amortized and MelT can be slower than the baseline ($\SpeedupOneAEighteen\times$--$\SpeedupOneMFour\times$ at 1~s on Apple Silicon); the advantage emerges with longer contexts or batched serving.
\end{itemize}

These boundaries follow from the mechanism quantified in Fig.~\ref{fig:roofline} rather than from implementation details, and they suggest a simple practice for pipeline optimization: profile the frontend rather than assume it negligible.

\section{Threats to Validity}
\label{sec:threats}

Several threats to validity are noted, together with the steps taken to mitigate them.

\subsection{Measurement and Instrumentation}

\emph{Software and hardware configuration.}
Realized latency and power depend on library versions, drivers, operating systems, compiler behavior, and backend implementations. These factors are reported in full (Table~\ref{tab:software}) to support reproducibility.

\emph{Backend arithmetic.}
The feature-equivalence analysis is computed in the same numerical environment as the speedup tables. CUDA and reference NumPy implementations were verified to agree within float32 rounding error, with maximum absolute differences below $2.4\times10^{-3}$ per operator.

\emph{Energy measurement.}
Energy is measured through hardware telemetry (NVML and \texttt{powermetrics}) rather than estimated from runtime alone. Because hardware counters are coarser than individual sub-millisecond kernel executions, power is interpreted as steady-state draw over repeated execution windows and energy values are reported as within-platform comparisons.

\emph{Edge-device proxy.}
The A18~Pro results are obtained on a MacBook~Neo carrying the same SoC as the
iPhone~16 Pro, used as a proxy for smartphone-class inference. The proxy is
\emph{fair} for the quantities reported here in three respects. First, the front-end kernels run on
physically identical compute units (shared microarchitecture, 3~nm process, and
unified-memory subsystem). Second, the headline edge metrics are within-device ratios
in which operating-system overhead, idle-power baseline, and chassis design act
on both front ends and largely cancel. Third, the workload is short-burst (a 160~s
clip completes in a few milliseconds, well below sustained-thermal limits). It
is \emph{necessary} because iOS exposes neither a full PyTorch/Metal runtime,
nor file-system access to the corpus, nor GPU power telemetry via
\texttt{powermetrics}, whereas the MacBook~Neo provides all three on identical
silicon. Accordingly the edge claims are scoped to per-call compute cost and
relative frontend comparison; these measurements do not reproduce sustained
device-level throughput under smartphone thermal constraints.

\emph{Profiling depth.}
Arithmetic-intensity values are analytical and kernel attribution is derived from PyTorch's operator-level profiler. Validation using lower-level vendor profiling tools such as Nsight Compute remains future work.

\subsection{Baseline and Comparison Fairness}

\emph{Baseline selection.}
For each platform, the fastest available implementation of every competing frontend was used. The reported speedups should be interpreted as comparisons against strong production baselines rather than against unoptimized reference implementations.

\emph{Baseline fusion.}
The conventional baseline and external libraries (TorchAudio, nnAudio, and \texttt{librosa}) represent production frontends rather than hand-written fused kernels. A compiler-fused baseline (\texttt{torch.compile}/TorchInductor) is reported in Section~\ref{subsec:extbaseline} and produces only modest changes because the conventional pipeline remains bandwidth-bound. Nevertheless, future work should evaluate fully hand-optimized fused STFT$+$Mel kernels. Such kernels would remain platform-specific---tied to a vendor FFT library such as cuFFT~\cite{nvidia_cufft} on NVIDIA or the Metal/Accelerate FFT on Apple---whereas the reformulated frontend is portable across all evaluated backends. Vendor data-loading pipelines such as NVIDIA DALI~\cite{nvidia_dali}, which expose GPU-accelerated spectrogram, Mel-filterbank, and MFCC operators, fall in the same category: their GPU path is CUDA-specific, so they accelerate the conventional pipeline within a single vendor's ecosystem rather than making it portable.

\subsection{Generality and Scope of Claims}

\emph{Equivalence margins.}
Equivalence margins are necessarily a modeling choice. The margins used in this work are derived from a parameter variation already regarded as negligible within the field (a 1\% $f_{\max}$ retuning) and are additionally cross-checked against fixed absolute margins.

\emph{Portability versus performance.}
The central claim of this work is portability of execution rather than universal superiority of performance. A GEMM-native frontend can execute on any accelerator exposing a matrix-multiplication primitive, but the magnitude of its latency and energy advantage remains dependent on hardware architecture, backend maturity, and frontend configuration. Consequently, portability does not imply identical speedups across devices.

\emph{Hardware-generation dependence.}
The experiments indicate that the realized benefits of the reformulation vary substantially across accelerator generations. Architectural characteristics such as memory hierarchy, GEMM throughput, and power efficiency influence the observed gains. Although the evaluated platforms span mobile, workstation, and datacenter settings, additional measurements on AMD, Intel, TPU, and future accelerator generations remain necessary.

\emph{Accelerator families.}
Evaluation covers NVIDIA (H100 and V100) and Apple Silicon (M4~Pro and A18~Pro). Behavior on other accelerator families, including AMD GPUs, Intel accelerators, and TPUs, remains uncharacterized.

\emph{Acoustic conditions.}
Equivalence is established on clean read speech (LibriSpeech) together with the clinical and speaker-attribute datasets used in this study. Robustness under severe environmental noise, channel distortion, reverberation, and other out-of-domain acoustic conditions is not evaluated.

\emph{Spectral resolution.}
The $\mathcal{O}(NM)$ projection loses its advantage as $M$ increases, approaching break-even near $M{=}512$ on the evaluated hardware. The compact-$M$ regime ($M\le128$) represents the intended operating region of the proposed formulation.

\emph{Model-capacity dependence.}
Drop-in equivalence holds for medium-sized and larger models. Smaller models (Whisper tiny/base) and a small time-resolved clinical CNN exhibit measurable degradation under zero-shot frontend substitution. A calibration layer may mitigate this behavior, but such approaches were not evaluated.

\begin{table}[!t]
\centering
\caption{\textbf{Software environment per platform, as logged during benchmarking}}
\label{tab:software}
\setlength{\tabcolsep}{6pt}
\resizebox{\columnwidth}{!}{
\begin{tabular}{lllll}
\toprule
Platform & OS & Python & PyTorch & GPU \\
\midrule
H100 & Linux (glibc 2.35) & 3.11.10 & 2.5.1+cu121 & H100 80GB HBM3 \\
V100 & Linux (glibc 2.17) & 3.8.19 & 2.3.1+cu121 & Tesla V100 32GB \\
M4 Pro & macOS 26.5.1 & 3.11.15 & 2.12.0 & --- \\
A18 Pro & macOS 26.5.1 & 3.11.15 & 2.12.0 & --- \\
\bottomrule
\end{tabular}
}
\end{table}

NVIDIA results used TF32-enabled CUDA builds of PyTorch (Tesla V100 driver 535.230.02). Apple results used the PyTorch MPS backend on the A18 Pro and the fastest available MLX GPU or MPS backend on the M4 Pro.

\section{Conclusion}
\label{sec:conclusion}

This work investigated whether a conventional audio frontend can be repositioned onto the matrix-multiplication substrate that already underlies modern accelerator execution. Using the well-established Non-Uniform Discrete Fourier Transform (NDFT), we reformulated Mel feature extraction as a GEMM-native computation and instantiated this principle through MelT and its cepstral extension MFCCT.

The resulting frontends demonstrate that such a reformulation is practical. Across heterogeneous accelerator platforms spanning mobile-edge, workstation, and datacenter deployments, the proposed formulation achieves up to a \SpeedupRangeHi$\times$ reduction in latency and a \EnergyMax$\times$ reduction in measured active energy (within-platform comparison). These computational gains do not require sacrificing downstream utility: word error rate remains statistically equivalent to the native frontend on frozen Whisper models of medium size and larger, speaker-attribute classification is non-inferior, and the reformulated cepstral frontend remains highly effective on a clinical respiratory-insufficiency classification task.

The NDFT-as-matmul identity itself is not claimed as a novel mathematical result. The contribution is instead a computational design principle. By decoupling the frontend from vendor-specific FFT primitives and expressing it directly through dense matrix multiplication, the frontend inherits the portability, optimization effort, and execution characteristics of the dominant computational substrate of modern accelerators. The results suggest that, in practical operating regimes, hardware alignment can outweigh arithmetic complexity as a predictor of realized latency and energy behavior.

More broadly, this work argues that audio frontends need not remain tied to FFT-centered execution models. MelT and MFCCT provide evidence that classical feature extraction can be reformulated around matrix-native execution while preserving practical utility. Exploring reformulations of other fixed-linear-transform frontends across additional signal-processing domains, accelerator families, and frontend architectures remains a direction for future research. As AI inference continues to expand from datacenter accelerators toward heterogeneous edge and on-device platforms, computationally portable, hardware-aligned frontends may become an increasingly relevant component of efficient inference systems \cite{tu2023energy}.
\section*{Acknowledgment}
During the preparation of this work, the author used Anthropic's Claude and Google's Gemini for two purposes: (i)~language editing, to improve the grammar, clarity, and readability of the manuscript text; and (ii)~refactoring, code review, and documentation of the accompanying open-source reproduction package. These tools were not used to generate research results, experimental measurements, or scientific claims. All experiments, analyses, and conclusions are the author's own; the author reviewed and verified all AI-assisted output and takes full responsibility for the content of this article.
\EOD

\bibliographystyle{IEEEtran}
\bibliography{main}

\end{document}